\documentstyle[aps,preprint,eqsecnum,epsf]{revtex}
\def\OP{\raisebox{.2ex}{$\stackrel{\leftrightarrow}{\bf P}$}}   
\def\OG{\raisebox{.2ex}{$\stackrel{\leftrightarrow}{\bf G}$}}   
\def\OS{\raisebox{.02ex}{$\stackrel{\leftrightarrow}{\bbox \Sigma}$}}  %
\begin{document}
\renewcommand{\thesection}{\arabic{section}}
\baselineskip=.8\baselineskip \count0=0 
{
\title{{\rm  LECTURE NOTES OF THE LES HOUCHES 1994 SUMMER SCHOOL}
\\
\bigskip
Exact  Resummations in the Theory of Hydrodynamic Turbulence:\\
0. Line-Resummed Diagrammatic Perturbation Approach}
\author {Victor L'vov\cite{lvov}  and Itamar  Procaccia\cite{procaccia}}
\address{Departments of~~\cite{lvov} Physics of Complex Systems
{\rm and}~~\cite{procaccia} Chemical Physics,\\
 The Weizmann Institute of Science,
Rehovot 76100, Israel,\\
\cite{lvov}Institute of Automation and Electrometry,
 Ac. Sci. of Russia, 630090, Novosibirsk, Russia
  }
\maketitle

 \widetext
\begin{abstract}
\baselineskip=.7\baselineskip 
\leftskip 40pt 
\rightskip 40pt 
  The lectures presented by one of us (IP) at the Les Houches summer
  school dealt with the scaling properties of high Reynolds number
 turbulence in fluid flows. The results presented are available in the
 literature and there is no real need to reproduce them here. Quite on the
 contrary, some of the basic tools of the field and theoretical techniques
 are not available in a pedagogical format, and it seems worthwhile to
 present them here for the benefit of the interested student. We begin
 with a detailed exposition of the naive perturbation theory for the
 ensemble averages of hydrodynamic observables (the mean velocity, the
response functions and the correlation functions). The effective expansion
parameter in such a theory is the Reynolds number (Re); one needs
therefore to perform infinite resummations to change the effective
expansion parameter. We present in detail the Dyson-Wyld line resummation
which allows one to dress the propagators, and to  change the effective
 expansion parameter from Re to O(1). Next we develop the ``dressed
vertex" representation of the diagrammatic series. Lastly we discuss in
full detail the path-integral formulation of the statistical theory of
 turbulence, and show that it is equivalent order by order to the
 Dyson-Wyld theory. On the basis of the material presented here one can
 proceed smoothly to read the recent developments in this field.

\end{abstract}
\leftskip 54.8pt

}
\narrowtext
\onecolumn   
\section{Introduction}
\label{20sect:intro}
The exact description of fluid flows is provided by the specification of
the velocity field ${\bf u}({\bf r},t)$ as a function of space and time.
Such a description is natural for ``laminar" flows of fluids whose
velocity field is not strongly fluctuating in space and time. On the other
hand, for ``turbulent" flows, which are states of fluids whose velocity
field ${\bf u}({\bf r},t)$ appears highly erratic as a function of ${\bf
r}$ and t, it is more sensible to consider a statistical description. The
statistical description is particularly useful in situations where the
flow is driven by stationary external agents. Examples of such stationary
driving mechanisms are found in shear turbulence, grid turbulence,
convective turbulence, etc. It is believed that in such flows it makes
sense to consider long time averages of functions of the velocity field
${\bf u}({\bf r},t)$.  For example, if the driving is arranged such that
the fluid as a whole is confined in some volume $V$, the mean velocity
must vanish:
\FL
\begin{equation}
\left\langle
{\bf u}({\bf r},t)
\right\rangle
\equiv{1\over V}\int d{\bf r}~
\lim_{T\rightarrow\infty}{1\over 2T}\int^T_{-T}{\bf u}({\bf r},t)dt=0\ .
\label{20a1}
\end{equation}
The pointed brackets indicate here an average with respect to
space and time. Likewise, correlation functions can be defined:
\begin{eqnarray}
&& \left\langle
{\bf u}({\bf r}+{\bf r}',t+\tau){\bf u}({\bf r},t)
\right\rangle
\label{20a2}   \\
&\equiv &
{1\over V}\int d{\bf r}\lim_{T\rightarrow\infty}{1\over
2T}\int^T_{-T}{\bf u}({\bf r}+{\bf r}',t+\tau){\bf u}({\bf r},  t)dt\ .
\nonumber
\end{eqnarray}
If the state of the fluid is statistically homogenous in space and time,
the averaging process (\ref{20a2}) results in a function of ${\bf r}'$ and
$\tau$ only.

One of the most important open problems of statistical physics is whether
the statistical description of turbulence displays universal
characteristics. Many important concepts were developed in the attempt to
answer this question. It was recognized by Richardson\cite{22Ric,26Ric} in
the `20's that quantities which depend on one space point have no chance
to claim universality. The velocity field ${\bf u}({\bf r},t)$ and all its
functions are fatally sensitive to the large scale components which are
introduced by the external driving agents. In this picture the driving
acts on the large scales, leading to maximally large velocity fluctuations
on the macroscale $L$, which is known as the ``outer" scale of turbulence.
The typical scale of the velocity fluctuations is denoted as $U_L$. The
order of magnitude of ${\bf u}({\bf r},t)$ at every ${\bf r}$ is about
$U_L$.  Richardson suggested that the quantities that may have universal
aspects are the velocity {\it differences} $\delta {\bf u} ( {\bf r}+ {\bf
R}, {\bf r},t) \equiv{\bf u} ( {\bf r}+ {\bf R},t) -{\bf u}( {\bf r},t)$.
In taking differences one eliminates the contribution of the large scale
motion, and may hope to see universal properties that are independent of
the driving mechanism. It was one of the main suggestions of Kolmogorov
that for $R$ that is much smaller than $L$ and much larger than a
``dissipative" scale $\eta$, statistical averages of functions of
 $\delta{\bf u}( {\bf r}+ {\bf R}, {\bf r},t)$ may show universality. The
dissipative scale $\eta$ depends on the kinematic viscosity of the fluid
$\nu$, and on the single most important dimensionless number in the theory
of turbulence, i.e.  the Reynolds number Re. This number is defined
as\cite{MY-2}
\begin{equation}
Re={U_LL\over \nu}\,,
\label{20a3}
\end{equation}
and in the Kolmogorov approach it is asserted that $\eta /L\sim
Re^{-3/4}$.  In other words, as Re$\rightarrow\infty$ there is a larger
and larger range of scales $R$ for which functions of $\delta{\bf u} (
{\bf r}+ {\bf R}, {\bf r})$ are expected to show universal properties.

In most experimental applications it is customary to consider the
longitudinal component of $\delta{\bf u}( {\bf r}+ {\bf R}, {\bf r},t)$
only:
\begin{equation}
\delta u( {\bf r}+ {\bf R}, {\bf r},t)
\equiv\delta{\bf u}( {\bf r}+ {\bf R}, {\bf r},t)\cdot{ {\bf R}\over R}\ .
\label{20a4}
\end{equation}
The essence of the Kolmogorov approach to the issue of universality in
turbulence, which was suggested in 1941\cite{41Kol-a} (K41) is that it is
possible to construct a local theory with one universal scaling exponent.
This universal exponent was ascribed to $\delta u( {\bf r}+ {\bf R}, {\bf
r})$, stating that
\begin{equation}
\delta u( {\bf r}+ {\bf R}, {\bf r})\sim R^{1/3}\ .
\label{20a5}
\end{equation}
It was asserted that $\delta u( {\bf r}+ {\bf R}, {\bf r})$ is the only
scaling field that has to be considered, in the sense that the structure
functions $S_n(R)$ which can be formed from $\delta u( {\bf r}+ {\bf R},
{\bf r},t)$ satisfy the scaling laws
\begin{eqnarray}
S_n(R)&\equiv &\left\langle\left\langle  [ \delta u(
{\bf r}+ {\bf R}, {\bf r},t)] ^n
\right\rangle \right\rangle
\nonumber \\
 & \sim & \big(\bar\varepsilon  R\big)^{n\zeta_n}
\sim(\bar\varepsilon R)^{n/3}
\label{20a6}
\end{eqnarray}
for values of $R$ in the ``inertial range" $L\gg R\gg\eta$.  In
(\ref{20a6}) $\bar\varepsilon$ is the mean (really in the double bracketed
sense of (\ref{20a1}) but defined by an overbar to save notation) of the
dissipation field $\varepsilon ( {\bf r},t)$,
\begin{equation}
\varepsilon ( {\bf r},t)\equiv{\nu\over 2}
\left[ \partial_\alpha u_\beta
( {\bf r},t)+ \partial_\beta u_\alpha ( {\bf r},t)\right]  ^2\ .
\label{20a7}
\end{equation}
This suggestion of Kolmogorov was immediately attacked both by theorists
(e.g. Landau) and experimentalists (for a recent review see
e.g.\cite{Fri}). Indeed, it is rather astonishing that a problem like
fluid turbulence, which suffers from very large fluctuations and strong
correlations, should be amenable to such a simple description; even
Kolmogorov himself revised his thinking and changed (\ref{20a6}) to a more
complicated form (which fell under attack as well).  Indeed, one
measurement that raised a lot of objections to the K41 approach is the
measurement of the correlation function of the dissipation field
$K_{\varepsilon\varepsilon}(r)$
\begin{equation}
K_{\varepsilon\varepsilon}(R)=\left\langle
\hat\varepsilon ( {\bf r}+{\bf R},t)\hat\varepsilon ( {\bf
r},t)\right\rangle\sim R^{-\mu}\ .
\label{20a8}
\end{equation}
where $\hat\varepsilon ( {\bf r},t)=\varepsilon ( {\bf  r},t)- \bar
\varepsilon$.  It was found in experiments that $K_{\varepsilon
\varepsilon}(R)$ decays very slowly in the inertial range, with $\mu$
having a numerical value in the range 0.20-0.25\cite{93SK}. It was claimed
that the K41 theory required $\mu$ to vanish. According
ly, there have been
many attempts to construct models of turbulence to take (\ref{20a8}) into
account and to explain how the measured deviations $(\zeta_n-1/3)$ in the
exponents of the structure functions  were related to $\mu$.

Our aim is to present a Navier-Stokes based theory of  turbulence.
This theory attempts to study the universality of of turbulence on scales
belonging to the bulk of the inertial range using the equations of fluid
mechanics.  We are not going to review the variety of phenomenological
{\it ad hoc} models which have proliferated in recent years.  It is our
belief that recent progress in the analytic theory brings us a long way
towards understanding the scaling properties of turbulence, and it is
important at this point to examine carefully what is known, to separate
false starts from genuine progress, and to delineate the domain of firm
results. In doing so we will also attempt to carefully assess the main
open problems that await further research.

The equations of fluid mechanics are partial differential equations, and
it is natural to consider them as a classical field theory, and to attempt
to apply techniques of perturbation theory and renormalization that found
spectacular success in the context of quantum field theory. Indeed, such
attempts were made with the help of various closure procedures like
Kraichnan's Direct Interaction Approximation\cite{59Kra} and many
others (for review of different closure schemes see,
e.g.\cite{70Ors}).  A more systematic  approach was suggested to be
found within diagrammatic perturbation approaches for non-equilibrium
processes like the Wyld diagrammatic technique\cite{61Wyl} for the
Navier-Stokes equation.  Generalizations of the Wyld diagrammatic
technique were suggested by Martin, Siggia, Rose \cite{73MSR} and by
Zakharov and L'vov\cite{75ZL}.

In performing such calculations one faces several serious difficulties.
Firstly, turbulence has no small parameter.  The naive perturbation
expansion of the Navier-Stokes equations results in increasing powers of
the Reynolds number, and this is rather inconvenient in a situation where
one is interested in the limit Re$\gg 1$.  Wyld\cite{61Wyl} found a
way to overcome this difficulty; by performing line renormalization like
Dyson's quantum field analog one finds a perturbative theory in which the
coupling parameter is of O(1).  This is an important step, but it does not
save Wyld's theory from severe problems. The main problem of the Wyld
theory is that it is written in terms of the correlation functions of the
field ${\bf u}({\bf r},t)$ itself. As mentioned above, such quantities are
not universal, and they are dominated by large scale contributions. In
addition, functions of ${\bf u}({\bf r},t)$ itself are not Galilean
invariant. Wyld attempted therefore to work in terms of Fourier transforms
of these quantities.  Indeed, the correlation functions in ${\bf
k},\omega$ representation are expected to be universal in the regime
$1/\eta\gg k\gg 1/L$.  Unfortunately, the theory in ${\bf k},\omega$
representation involves integrals over the whole ${\bf k},\omega$ range,
and the theory picks up infra-red divergences that cannot be eliminated
easily.

Kraichnan has suggested that a theory that does not suffer from the
problems of Wyld's expansion can be formulated in Lagrangian
coordinates\cite{65Kra,66Kra}.  From the point of view of developing the
theory in terms of universal quantities this suggestion was correct.
Unfortunately the perturbation theory which is based on the Lagrangian
representation is not of the diagrammatic type, and higher order terms
with its coefficients cannot be found simply on the basis of
topological features of the diagrams. We shall not be interested here
in theories that contain arbitrary truncations.  Such theories are
uncontrolled, and it is our hope that one can achieve a fully controlled
theory of turbulence. We seek therefore a formulation that lends itself to
a diagrammatic description and which comes as close as possible to using
as building blocks functions of $\delta u( {\bf r}+ {\bf R}, {\bf r})$.  A
theory that has all the right ingredients was formulated by Belinicher and
L'vov\cite{87BL} (see also\cite{91Lvo}).  This theory went under the name
``quasi-Lagrangian"; this name turned out to be unfortunate, since it left
the impression of an approximate theory.  Recently \cite{95LP-b} we
revisited this theory and reiterated that it is not an approximate theory,
but rather an exact renormalization scheme that differs from the Wyld
technique.

The aim of these notes is to review the basic tools of the field theoretic
diagrammatic approach to the Navier-Stokes problem. There are essentially
no new developments in these notes, but they are meant to bring the
student to a level of expertise that would allow a smooth entry into the
latest developments. We develop in full detail the naive perturbation
theory (sec.2) for the Green's function and the 2-point correlation
function. This theory is badly diverging in powers of Re. In sec.3 we
discuss the line resummation which results in the Dyson equation for the
Green's function and the Wyld equation for the correlator. In these
equations, which are also presented as infinite diagrammatic series
expansions, the effective coupling constant is of O(1). We discuss briefly
3-point and higher order correlation functions, and in Sect.~4 we turn to
the functional integral formulation of the renormalized perturbation
theory. We show in detail that the diagrammatic expansion obtained in
this formulation is identical, order by order, with the Wyld
formulation, and that either one can be used at
will\cite{73MSR,76Dom,76Jan}.  The interested student is invited then to
continue reading Refs.  \cite{95LP-b,95LP-c,95LP} in which the
Belinicher-L'vov renormalization scheme is used to develop a consistent
theory of normal and anomalous scaling in turbulence.

\section{Naive Perturbation Theory}
\label{20sect:naive}
\subsection{Introduction}
\label{20sect:naive-1}

As was mentioned in section \ref{20sect:intro}, we are mostly interested
in the calculation of averages and correlation functions, which are
obtained in the long time limit of the {\it forced} Navier-Stokes
equations.  Since the equations contain a viscous damping, it is assumed
that the balance between forcing and damping leads after a time to
stationary statistics which are independent of the initial conditions.
Roughly  speaking the picture is of a nonlinear problem that generates its
own ergodic measure, and the forcing is chosen such that the
ergodic measure (at least as far as the calculated quantities are
concerned) is a property of the Navier-Stokes equations and not of the
forcing. In other words, the solution roams forever on a strange
attractor, and the forcing is used mainly to keep the system away from the
(stable) zero solution, but without ruining too much the measure on the
attractor.

In our formal scheme we will expand ergodic averages for the nonlinear
problem in terms of averages of solutions of the linear problem (i.e with
the nonlinearity discarded). Not surprisingly, since the measure of the
nonlinear problem is {\it very} different from the measure of the linear
problem, we will see that our expansion has very miserable convergence
properties. Actually the naive expansion is diverges, and it cannot be
truncated at any order. Its has meaning only for the series as a whole.
Nevertheless, one can perform infinite partial resummations of this series
to achieve a new series that has somewhat better properties, i.e. the
terms remain all of the same order. We stress right at the beginning that
many of our steps have dubious mathematical validity, and the fact that
they are common practice in many field theories cannot be really taken as
an excuse. At least we will not hide the difficulties.

The Eulerian velocity field is denoted in our discussion as ${\bf u}({\bf
r},t)$.  In later notation we shall sometimes use the 4-vector notation
${\bf u}(x)$, where $x\equiv ({\bf r},t)$. In the mathematical literature
it is customary to write the Navier-Stokes equations as an initial value
problem without any type of forcing:
\FL
\begin{equation}
{\partial{\bf u}\over   \partial t}+({\bf u}\cdot{\bbox{\nabla}} )
{\bf u}-\nu{\nabla}^2{\bf u}-{\bbox\nabla} p=0\,,\ \
{\bbox\nabla}\cdot{\bf u}=0\ .
\label{20b1}
\end{equation}
For the physicist the equations of fluid mechanics are a projection onto a
set of slow variables of the many-body microscopic dynamics. In such a
projection, when successful, the host of irrelevant degrees of freedom
remain in the form of a ``noise" term in the equation of motion of the
slow variables. The basic assertion is that the irrelevant degrees of
freedom are much faster, and reach local equilibrium on a short time scale
which is irrelevant for the hydrodynamic description.  Thus, the physicist
writes Eq. (\ref{20b1}) in the form
\begin{equation}
\partial{\bf u}/   \partial t
+({\bf u}\cdot{\bbox\nabla} ){\bf u}-\nu{\nabla}^2{\bf u}
-{\bbox\nabla} p={\bf f},
\label{20b2}
\end{equation}
where the fluctutation dissipation theorem dictates that the two-point
correlation function of the ``random" force ${\bf f}$ is related to the
viscosity and the temperature of the fluid:
\begin{equation}
\overline{f_\alpha ({\bf r},t)f_\beta ({\bf r}',t')}
=2\nu T\delta ({\bf r}-{\bf r}')\delta (t-t') \delta_{\alpha\beta}\ .
\label{20b3}
\end{equation}
An overbar denotes an average with respect to the thermodynamic
equilibrium ensemble with  temperature $T$.

In natural conditions turbulence arises either due to the instabilitiy of
laminar flows (like in channel flow) or due to time dependent stirring
(such as in a glass of water stirred by a spoon). For the sake of
theoretical simplicity we will model all these situations with a stirring
force ${\bbox\phi} ({\bf r},t)$. The properties of the turbulent flow
 may depend on the nature of the stirring force ${\bbox\phi} ({\bf r},t)$.
 Since in experimental situations one finds that the statistics of the
turbulent velocity field on the large scales are close to that of a
Gaussian random ensemble, it is customary to take ${\bf f} ({\bf r},t)$ to
be a Gaussian random force with spectral support in the small wave
vectors and small frequencies. The properties of the correlation function
of ${\bbox\phi} ({\bf r},t)$ are best stated in ${\bf k},\omega$ space:
it is concentrated in the small $k,\omega$ region, i.e.  $k\leq 1/L$,
$\omega\leq U_{_L}$ where $U_{_L}$ is the characteristic magnitude of
turbulent velocity fluctuations. In order to model properly the generic
properties of turbulent fluids one needs to assume that the correlation
decays sufficently quickly to zero for $k\gg 1/L,~\omega\gg U _{_L}/L$.
In ${\bf r},t$ space it means that $D(R,\tau )$,
\begin{equation}
D_{\alpha\beta}(R,\tau)=\left\langle\left\langle
\phi_\alpha ({\bf r}+{\bf R},t+\tau
)\phi_\beta({\bf r},t)\right\rangle\right\rangle
\label{20b4}
\end{equation}
is constant for $r\ll L,~\tau\ll L/U$ and for higher values of $R$ and
$\tau$ it decays quickly. One of the important questions that the theory
should address is whether the properties of turbulence on scales much
smaller than $L$ are independent of the precise choice of $D(R,\tau )$.
Also, the possible effect of non-Gaussianity of the forcing should be
understood.

Adding the stirring force to the equation of motion we write
\begin{equation}
\partial {\bf u}/   \partial t
+({\bf u}\cdot{\bbox\nabla} ){\bf u}-\nu{\nabla}^2{\bf u}
-{\bbox\nabla} p= {\bf f}+{\bbox\phi}\ .
\label{20b5}
\end{equation}
The combination ${\bf f}  + {\bbox\phi}$ will be referred to, when
convenient, as $\tilde {\bf f}$.

In this work we deal with incompressible turbulence only, in which
${\bbox\nabla}\cdot{\bf u}=0$. We shall therefore project out from Eq.
(\ref{20b5}) any longitudinal components. This is done with the help of
the projection operator $\OP$
which is formally written as
${\OP}
\equiv -{\nabla}^{-2}{\bbox\nabla} \times{\bbox\nabla} \times$.  In tensor
notation this operator is $P_{\alpha\beta}$ $=\delta_
{\alpha\beta}-{\nabla}^{-2} {\bbox\nabla}_\alpha{\bbox\nabla}_
\beta$.  Because of the inverse Laplace operator which is nonlocal in
space, the application of $\OP$
to any given vector field ${\bf a}({\bf r})$ is non local:
\begin{equation}
[ \OP{\bf a} ({\bf r})] _\alpha
=\int d{\bf r}' P_{\alpha\beta} ({\bf r}-{\bf r}')a_\beta
({\bf r}')\,,
\label{20b6}
\end{equation}
where $P_{\alpha\beta}({\bf r}-{\bf r}')$ is the inverse Fourier
transform of $P_{\alpha\beta}({\bf k})$
\begin{equation}
P_{\alpha\beta}({\bf r}-{\bf r}')=\int{d{\bf k}\over (2\pi)^3}
\exp [-i({\bf r}-{\bf r}')\cdot{\bf k} ] P_{\alpha\beta}({\bf k}).
\label{20b7}
\end{equation}
The tensor $P_{\alpha\beta}({\bf k})$ in $k$-representation is
\begin{equation}
P_{\alpha\beta}({\bf k})=\delta_{\alpha\beta}
-{1\over k^2}k_\alpha k_\beta
\label{20b8}
\end{equation}
and in $r$-representation
\begin{equation}
P_{\alpha\beta}({\bf r}-{\bf r}')
={\delta_{\alpha\beta}\over\vert{\bf r}-{\bf r}'\vert^3}
-3 {(r_\alpha-r'_\alpha)(r_\beta-r'_\beta)
\over\vert{\bf r}-{\bf r}'\vert^5} \ .
\label{20b9}
\end{equation}
The operator $\OP$
has the following properties:  (i)
${\bbox\nabla}\cdot\OP$
$ =\OP \cdot{\bbox\nabla} =0$.
(ii)  For any divergence-free vector $\bf b$,
$\OP {\bf b}={\bf b}$.
Applying $\OP$
to Eq. (\ref{20b5}) we find
\begin{equation}
(   \partial/   \partial t-\nu{\nabla}^2){\bf u}+
\OP
({\bf u}\cdot {\bbox\nabla} ){\bf u}=\OP
\tilde  {\bf f}\ .
\label{20b10}
\end{equation}
Introduce the bare Green's operator ${\bf G}_0$ via the relationship
\begin{equation}
\OG_0
\equiv{i\over   \partial/ \partial t-\nu{\nabla}^2}
\OP\ .
\label{20b11}
\end{equation}
Using $\OG_0$ we can rewrite Eq. (\ref{20b10}) in the form:
\begin{equation}
{\bf u}({\bf r},t)
={\bf u}_0({\bf r},t)-i\OG_0
({\bf u}\cdot{\bbox\nabla} ){\bf u}\ ,
 \label{20b12}
\end{equation}
where
\begin{equation}
{\bf u}_0({\bf r},t)\equiv -i
\OG
_0\tilde {\bf f}\ .
\label{20b13}\end{equation}
The meaning of the application of $\OG_0$ to
any vector field ${\bf c}( {\bf r},t)$ is understood by
\begin{eqnarray}
\OG_0
{\bf c}( {\bf r},t)
&=&G^0_{\alpha\beta}({\bf r}-{\bf r}',t-t') *{\bf c}
({\bf r}',t)
\label{20b14} \\
&\equiv&\int d{\bf r}'dt'
G^0_{\alpha\beta}({\bf r}-{\bf r}',t-t')
c_\beta ({\bf r}',t')\,,
\nonumber
\end{eqnarray}
where the kernel $G^0_{\alpha\beta}({\bf r}-{\bf r}',t-t')$ is the
Green's function of the linear part of Eq. (\ref{20b10}), and we have
introduced the * operation according to the RHS of Eq. (\ref{20b14}). The
explicit representaion of $G^0_{\alpha\beta}({\bf r}-{\bf r}',t-t')$ is
given as the inverse transform in {\bf k} and $\omega$
\begin{eqnarray}
&&G^0_{\alpha\beta}({\bf r}-{\bf r}',t-t')
\label{20b15}    \\
&=&\int{d{\bf k}
d\omega\over (2\pi)^4}
\exp i[ {\bf k}\cdot ({\bf r}-{\bf r}')+\omega
(t-t')] G^0_{\alpha\beta}({\bf k},\omega )\ .
\nonumber
\end{eqnarray}
The function $G^0_{\alpha\beta}({\bf k},\omega )$ is
\begin{equation}
G^0_{\alpha\beta}({\bf k},\omega )={P_{\alpha\beta}({\bf
k})\over \omega +i \nu k^2}\ .
\label{20b16}
\end{equation}
The convention chosen is that $G^0_{\alpha\beta}({\bf k},\omega )$ is
analytic in the upper half of the plane of complex $\omega$.
Correspondingly $G^0_{\alpha\beta}({\bf r}-{\bf r}',t-t')$ is zero for
$t'< t$.  This property is known as causality, and it stems from the
phyiscal constraint that a response cannot come before its cause. It is
obvious that the solution (\ref{20b12}) involves integrals over all space
in  ${\bf r},t$ representation because of the inverse operators in $\OG_0$
which is a factor function in ${\bf k},\omega$ space. The solution
involves integrals also in ${\bf k},\omega$ space due to the nonlinear
term which is a local operator in ${\bf r},t$ space.  Thus there is no
formal advantage to either representation, and we shall write Eq.
(\ref{20b12}) in the representation invariant form
\begin{equation}
{\bf u}={\bf u}_0+{1\over  2}
{\bf G}^0*\Gamma ^{\cdot{\bf u}}_{\cdot{\bf u}}  \ .
\label{20b17}
\end{equation}
The~~*~~and the ~$\cdot$~ operations depend on the
representation. In ${\bf r},t$ representation the ~*~ operation was
defined by Eq.  (\ref{20b14}), and $\Gamma ^{\cdot{\bf u}}_{\cdot{\bf u}}$
means $2({\bf u}\cdot{\bbox\nabla} ){\bf u}$.  For later applications we
need to consider the operation $\Gamma ^{\cdot{\bf u}}_{\cdot{\bf u}}$ in
which different fields {\bf u} and $\tilde {\bf u}$ are involved. We will
interpret $\Gamma ^{\cdot{\bf u}}_{\cdot\tilde {\bf u}}$ in a symmetric
fashion, i.e.

\begin{equation}
\Gamma ^{\cdot{\bf u}}_{\cdot\tilde {\bf u}}
=-i[  ({\bf u}\cdot{\bbox\nabla} )\tilde{\bf u}+ (\tilde{\bf u}
\cdot{\bbox\nabla} ){\bf u}]\ .
\label{20b18}
\end{equation}
In Fourier representation we write
\begin{eqnarray}
{\bf u}({\bf k},\omega )&\equiv&
\int d{\bf r}  dt \exp[-i({\bf r}\cdot{\bf k}+\omega t)]
{\bf u}({\bf r},t)\,,
\label{20b19} \\
\left[{\bf G}^0*{\bf c}\right]_\alpha&=&G^0_{\alpha\beta} ({\bf k},\omega )
c_\beta ({\bf k},\omega )\,,
\label{20b20}
\end{eqnarray}
 and
\begin{eqnarray}
&&[ \Gamma^{\cdot{\bf u}}_{\cdot\tilde {\bf u}}]
 _\alpha ({\bf k},\omega)
= \int{d {\bf k}'\over  (2\pi )^3}{d{\bf k}''\over  (2\pi )^3}
{d\omega'\over  2\pi} { d\omega''\over 2\pi }
\nonumber\\
&\times&\Gamma_{\alpha\beta\gamma} (q,q',q'')u^*_\beta
({\bf k}',\omega ')\tilde u^*_\gamma ( {\bf k}'',
\omega '')
\label{20b21}
\end{eqnarray}
where
\begin{eqnarray}
\Gamma_{\alpha\beta\gamma} (q,q',q'')&=&
(2\pi )^4(k_\beta\delta_{\alpha\gamma}
+ k_\gamma\delta_{\alpha\beta} )
\nonumber\\
&\times&\delta ({\bf k}+{\bf k}'+{\bf k}'')
\delta (\omega +\omega'  +\omega'')\ .
\label{20b22}
\end{eqnarray}
In these equations and below we introduced the four dimensional vector
$q\equiv \{{\bf k},\omega \}$.  The symmetry of (\ref{20b21}) with respect
to exchanging ${\bf u}$ and $\tilde {\bf u}$ is inherited from
(\ref{20b22}).

\subsection{Naive Perturbation Theory for the velocity field}
\label{20sect:naive-2}
It is natural to seek solutions of Eq. (\ref{20b17}) in the form
\begin{equation}
{\bf u}({\bf r},t)=\sum^\infty_{n=0}{\bf u}_n,~~~{\bf u}_n
\propto\Gamma^n     \ .
\label{20c1}
\end{equation}
Substituting this in the Navier-Stokes equation in the form
(\ref{20b17}), and collecting terms with the same power in $\Gamma$ we
have a set of recurrence relations starting with \begin{eqnarray} {\bf
u}_1&=& {1\over 2}{\bf G}^0*\Gamma^{\cdot{\bf u}_0}_{\cdot{\bf u}_0}\,,
\label{20c2}\\
{\bf u}_2&=&{1\over
2}{\bf G}^0*\Gamma^{\cdot{\bf u}_1}_{\cdot{\bf u}_0} + {1\over 2}
{\bf G}^0*\Gamma^{\cdot{\bf u}_0}_{\cdot{\bf u}_1}
={\bf G}^0*\Gamma^{\cdot{\bf u}_1}_{\cdot{\bf u}_0}\ .
\label{20c3}
\end{eqnarray}
The last equality follows from the symmetry of $\Gamma$ which was
discussed in the previous section. In fact, we are making full use here of
the fact that we study a classical system. In the corresponding quantum
mechanical theory the two terms in Eq. (\ref{20c3}) are distinct due to
the non-commutativity of ${\bf u}_1$ and ${\bf u}_0$ which become
operators.

The next orders have the form
\begin{eqnarray}
{\bf u}_3&=&{\bf G}^0*\Gamma^{\cdot{\bf
u}_2}_{\cdot{\bf u}_0}+{1\over 2} {\bf G}^0*\Gamma^{\cdot{\bf
u}_1}_{\cdot{\bf u}_1}\,,
\label{20c4}\\
{\bf u}_4&=&{\bf G}^0*\Gamma^{\cdot{\bf u}_3}_{\cdot{\bf u}_0}+ {\bf
G}^0*\Gamma^{\cdot{\bf u}_2}_{\cdot{\bf u}_1}\,,
\label{20c5}
\end{eqnarray}
etc. In general,
\begin{eqnarray}
{\bf u}_n&=&
\sum^{n/2-1}_{m=0}{\bf G}^0*\Gamma^{\cdot {\bf u}_{n-m+1}}_
{\cdot{\bf u}_m}~~~(n~{\rm even})\,,
\label{20c6}\\
{\bf u}_n&=&\sum^{(n-3)/2}_{m=0}{\bf G}^0*\Gamma
^{\cdot {\bf u}_{n-m+1}}_{\cdot{\bf u}_m}
\nonumber\\
&&+ {1\over 2}{\bf G} ^0*\Gamma
^{\cdot{\bf  u}_{(n-1)/2}}_{\cdot{\bf  u}_{(n-1)/2}}~~(n~{\rm odd})\ .
\label{20c7}
\end{eqnarray}
At this point one may substitute lower order ${\bf u}_m$ in (\ref{20c6})
and (\ref{20c7}) to achieve eventually a solution in terms of ${\bf u}_0$
and ${\bf G}^0$. For example ${\bf u}_2$ can be written as
\begin{equation}
{\bf u}_2={\bf G}^0*\Gamma
^{\cdot (1/2)\Gamma^{\cdot{\bf u}_0}_{\cdot{\bf u}_0}}_{\cdot{\bf u}_0}
\label{20c8}
\end{equation}
\begin{figure}
\epsfxsize=6truecm
\centerline{\epsfbox{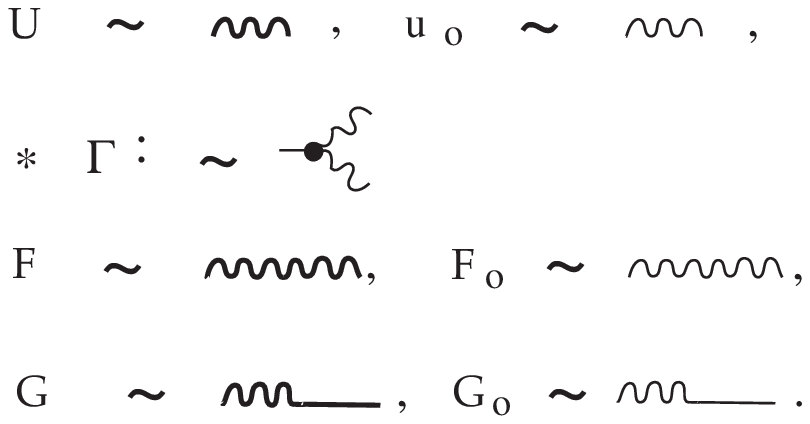}} 
\caption{
Graphical notation for the perturabtion expansion.  The symbol  used are
the following: short wavy lines stand for the fluid velocity  (think about
waves). A thin wavy line stands for ${\bf u}_0$, whereas a bold  wavy line
represents the full solution ${\bf u}({\bf r},t)$. A straight line stands
for the field ${\bf p}({\bf r},t)$ that is only introduced in Chapter
III.  The Green's function, which is the response in the velocity to some
force is made of a short wavy line and a short straight line representing
the force.  Again this symbol appears in thin and bold variants. The
former stands for the bare Green's function (2.11), and the bold for the
dressed Green's function (3.5). The vertex  (2.18) is a fat dot with three
tails.  One straight tail belongs to the Green's function, and two wavy
tails stand for velocities.  A long wavy line will represent correlation
functions of velocities.  Again, the thin and bold variants are bare and
dressed correlators respectively. }
\label{20-fig1}
\end{figure}
For higher values of $n$ this substitution becomes increasingly more
cumbersome in analytic form, and it is very helpful to represent it
graphically.  The notation used is shown in Fig.~\ref{20-fig1}. In
Fig.~\ref{20-fig2}a
we display the diagrammatic representation of the equation of motion, and
in Fig.~\ref{20-fig2}b and Fig.~\ref{20-fig2}c we show the diagrammatic
form of Eqs.  (\ref{20c2}) and (\ref{20c8}). Examining the other graphs in
Fig.~\ref{20-fig2} we see the great advantage of the diagrammatic
representation. It is very easy to write the n'th order diagrams without
going through the cumbersome analytic substitutions. The rules for drawing
the diagrams are very simple: for the n'th order term we draw all the
topologically distinct binary trees with $n$ vertices, such that all the
trunks are made of Green's functions and all the end branches are made of
${\bf u}_0$'s.  Every portion of the tree that continues in a symmetric
fashion gets a factor of 1/2. That is all.  The existence of these simple
rules for finding the $n$'th order term, which are independent of $n$,
reflects the internal sturcture of this perturabtion theory, and it will
allow us to perform resummations in a straightforward way.

It should be understood that the perturbation theory described in this
section is not expected to converge. Consider for example the ratio of
${\bf u}_1$ over ${\bf u}_0$. ``Dividing" Eq. (\ref{20c2}) by $u_0$ leaves
us with the order of magnitude estimate
\begin{equation}
{\bf u}_1/{\bf u}_0
\sim{\bf G}^0*\Gamma^{\cdot{\bf u}_0}_\cdot\sim
{\bbox\nabla}{\bf u}_0/\nu{\nabla}^2\sim
{\bf u}_0L/\nu={\rm Re}\ .
\label{20c9}
\end{equation}
We see that we are effectively expanding in Reynolds number, and for high
Re this is a terribly divergent series. The main problem in this
expansion is the appearance of the molecular viscosity $\nu$ in the
denominator in (\ref{20c9}). It is expected that in a turbulent fluid the
viscosity is strongly renormalized because of the interaction, yielding
the so called ``eddy-viscosity".  We need to reformulate the theory such
that the eddy-viscosity appears in the effective expansion parameter to
make it of order 1. This is achieved in Sect.\ref{20sect:naive}.

\begin{figure}
\epsfxsize=8.6truecm
\centerline{\epsfbox{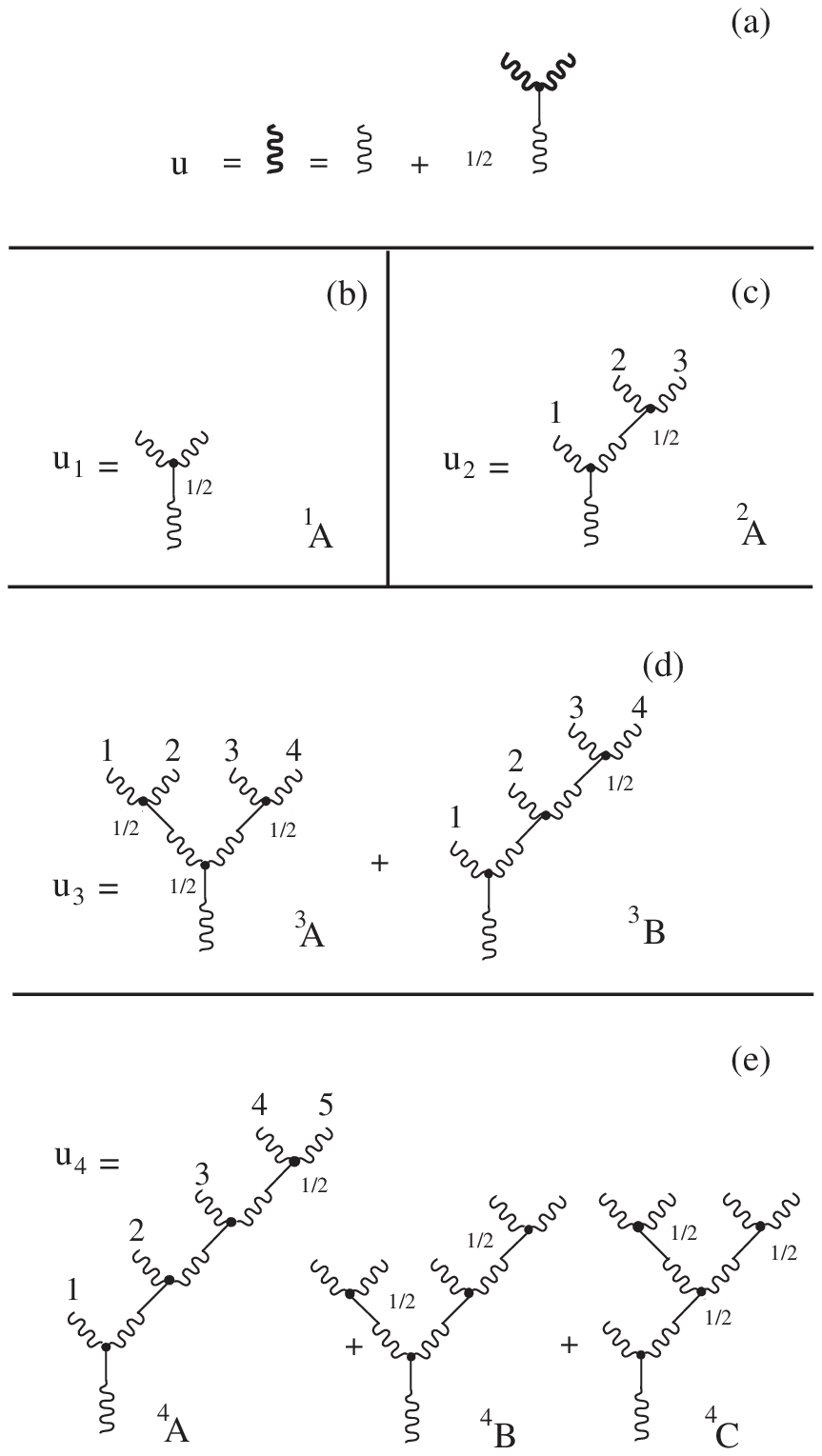}} 
\caption{
Graphic representation of the equation of motion [Panel (a)] and the
naive perturbation theory. The diagrams in Panels (b)-(d) are all the
contributions appearing in ${\bf u}_1-{\bf u}_4$. Note that all the
topologically distinct trees appear. These trees  have a trunk that begins
with the wavy line of a bare Green's function, and continues all the way
via bare Green's functions. The branches are made of wavy lines of ${\bf
u}_0$.  The bifurcations carry a vertex $\Gamma$. The factors of 1/2
appear in all the  bifurcations that have symmetry above them. The total
factor in front of a tree is the product of these numbers. Every tree has
been denoted with a letter and a superscript. The latter stands for the
order in $\Gamma^n$.  For future purposes we have numbered the branches of
the trees $^2A, ^3A, ^3B$ and $^4A$.   }
\label{20-fig2}
\end{figure}

\subsection{Naive perturbation theory for statistical quantities}
\label{20sect:naive-3}
\subsubsection{Statistics}

As explained in the Introduction, our aim is to formulate a theory for
ergodic averages. In the naive perturbation theory we can compute such
averages by using the expansion (\ref{20c1}) for ${\bf u}({\bf r},t)$, and
averaging products of ${\bf u}_0({\bf r},t)$ at different points and times
with respect to different realizations of the random forcing  $\tilde {\bf
f} $. In doing so we shall make full use of the fact that $\tilde {\bf f}
$ is assumed to be Gaussian. On the one hand, this leads  to a great
simplification of the naive perturbation theory. On the other hand it
seems theoretically dangerous. The physics that we are trying to describe
is surely non-Gaussian. Are we not throwing away important aspects of the
physics by using from the start a Gaussian forcing? The answer will be NO.
We shall see later that the structure of the theory in its final resummed
formualtion does not depend on this way of constructing the naive
expansion. The physics is dominated by strong interactions via the
nonlinearity, and a Gaussian forcing is sufficient to excite all the
possible responses which the dynamics amplifies and redistributes to allow
us to observe the full statistical structure. In fact, at the end of the
calculation we are going to explore the limit in which the random forcing
becomes vanishingly small. We shall show that this limit exists, and that
in that limit the statistics of velocity fluctuations becomes independent
of the statistics of the random force. Notwithstanding, it should be said
here that it is not at all obvious that the procedure followed here is the
most {\it efficient} one. It is possible that one can get a better
converging theory by forcing the system with a force that is closer in
character to the statistics of the dressed system. We do not know how to
do that, and this is one of the open problems that needs to be studied in
the future. We shall return to this issue in Sec. 3 after the formal
apparatus needed for its full appreciation is already at hand.

We will denote averages with respect to the realization of the random
force with single pointed brackets, in contrast with the double pointed
brackets of Eq. (\ref{20a1}). The Gaussianity of the random force means
that
\begin{eqnarray}
\langle{\tilde {\bf f}} (x)\rangle &=&0\,,
\nonumber\\
\langle{\tilde {\bf f}} (x_1){\tilde{\bf f}} (x_2)\rangle &=& D_{12} \,,
\nonumber\\
\langle{\tilde{\bf f}} (x_1)
{\tilde{\bf f}} (x_2)
{\tilde{\bf f}} (x_3)\rangle &=&0
 \label{20d1}\\
\langle
{\tilde{\bf f}} (x_1){\tilde{\bf f}} (x_2){\tilde{\bf f} }
(x_3){\tilde{\bf f}} (x_4) \rangle
&=&D_{12}D_{34}+D_{13}D_{24}+D_{13}D_{23}\,,
\nonumber
\end{eqnarray}
etc.  All higher order correlations with odd number $\tilde{\bf f} $
vanish, and the correlations involving $2n$ factors of $\tilde{\bf f} $
have $(2n-1)!!$  contirbutions corresponding to all the possible pairings
of $\tilde{\bf f} $.
\begin{figure}
\epsfxsize=8.6truecm
\centerline{\epsfbox{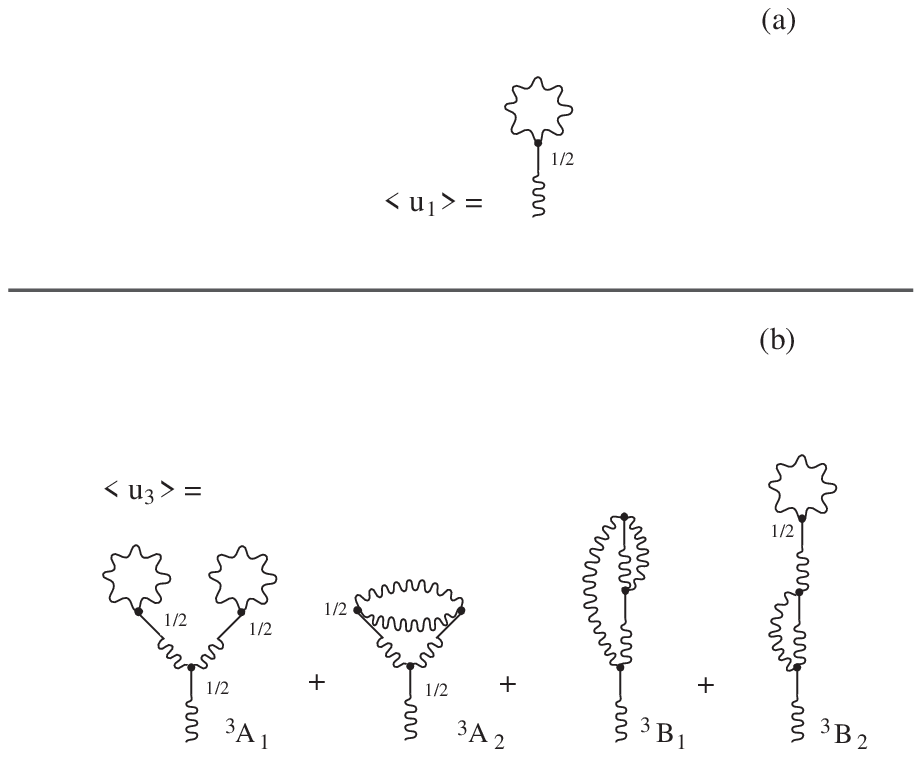} } 
\caption{
The diagrams representing the mean velocity. Panel (a) represents the
first order contribution in {$\Gamma$}, and is obtained from the simple
gluing of the two branches of the tree {$^1A$} in Fig.~2b. It carries the
factor of 1/2 which originates from the symmetry of {$^1A$}. Clearly,
averaging either {$^2A$}  of Fig.~2c or any tree in Fig 2e results in a
zero contribution due to the  odd number of branches which carry a random
force:  {$\langle{\bf u}_2\rangle =\langle{\bf u}_4\rangle =0$}.  Panel
(b) represents the result of averaging of the trees {$^3A$}  and {$^3B$}
in Fig.~2d.  The diagran {$^3A_1$} originates from gluing of the branches
1 with 2 and 3  with 4 in {$^3A$}.  Diagram {$^3A_2$} comes from gluing
either 2 with 3 and 1 with 4 or 1 with 3 and 2 with 4 in {$^3A$}.
Correspondngly we gain a factor of 2 with {$^3A_2$} leading to a
coefficient 1/4 instead of 1/8 in {$^3A$}. The tree {$^3B_1$} originates
from {$^3B$} after gluing 1 with 3 and 2 with 4 or 1 with 4  and 2 with 3.
A factor of 2 is gained. Lastly, diagram {$^3B_2$} results from  gluing 1
with 2 and 3 with 4 in {$^3B$}. There is only one way of doing it,  and
the factor 1/2 remains.  The general rule for the overall factor in
front of a diagram is obtained as follows: count the number of vertices
such that that exchanging the two branches emanating from them  leaves the
diagram invariant. Denote the numebr of distinct pairs of  such branches
by {$N$}. The overall factor in front of the diagram is {$1/2^N$}.
}
\label{20-fig3}
\end{figure}

\subsubsection{The mean velocity}
Of the sought statistical quantities, the easiest to obtain is the mean
velocity, averaged over all the possible realizations of the random
force. Since this random force is Gaussian, we have well defined
statistics for the averaging process. We can apply the rules (\ref{20d1})
to the average of the diagrammatic representation of ${\bf u}({\bf r},t)$.
We pair the ${\bf u}_0$ branches in all the possible ways, and glue the
ends together, see Fig.~\ref{20-fig3}. Every diagram with an odd number of
${\bf u}_0$ branches gives no contribution. Every diagram with 2n ${\bf
u}_0$ branches gives (2n- 1)!! contributions which are obtained from all
the possible binary pairings of random forces. The process is shown in
Fig.~\ref{20-fig3}. The diagrams $^nA_m$ and $^nB_m$ for $\langle{\bf
u}_n\rangle$ in Fig.~\ref{20-fig3} result from the diagrams $^nA$ and
$^nB$ for ${\bf u}_n$ in Fig.~\ref{20-fig2}. Note that in systems which
are homogeneous and isotropic the mean velocity vanishes.  Consequently,
the sum of all the diagrams obtained in this fashion has to vanish. This
will be used in our later developments. In a turbulent system with a space
dependent mean velocity profile this set of diagrams will not vanish, and
it will contribute also in other statitsitcal averages that we consider
below.  These diagrams will describe the interaction of the mean profile
with the velocity fluctuations.

\subsubsection{The Green's function}
Next we discuss the Green's function, which is the response of the
velocity field to an external perturbation.  The Green's function is
defined as
\begin{equation}
G_{\alpha\beta}(x,x')=i\left\langle\delta u_\alpha(x)/
\delta{\tilde f}_\beta (x')\right\rangle
\label{20d2}
\end{equation}
where the notation $\delta(\cdot )/\delta (\cdot )$ stands for
the functional derivative. The meaning of this functional derivative is
the following:  solve the Navier-Stokes equations once with a forcing
$\tilde{\bf f} $ and once with a forcing $\tilde{\bf
f}+{\bbox\epsilon}\delta (x-x')$.  Then take the ratio $[ {\bf
u}(x,{\bbox\epsilon})-{\bf u}(x,0)]  / {\bbox\epsilon}$ in the limit
$\epsilon\rightarrow 0$, and average over the ralizations of the random
force. The principle of causality means that ${\bf G}(x,x')$ is zero for
$t'< t$.  This property will be used a lot in the sequel.

\begin{figure}
             \epsfxsize=8.6truecm
\centerline{             \epsfbox{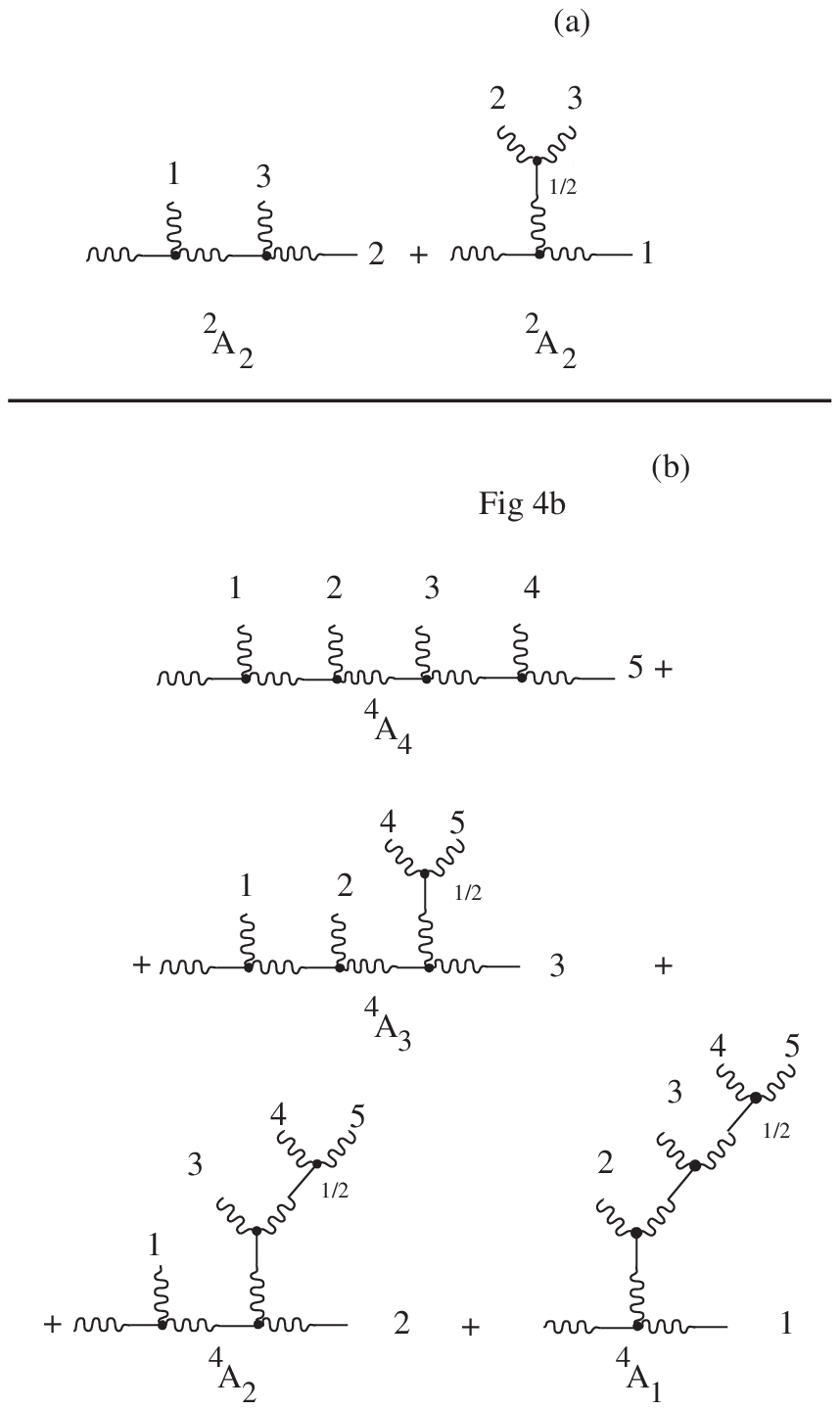} } 
             \vspace{.5cm}
\caption{
Typical contribution to the diagrammatic representation of
{$\delta{\bf u}(x)/\delta\tilde{{\bf f} }$} which originates from the
trees $^2A$ and $^4A$ in Fig.~2c and  2e. In panel (a) we show
the diagrams originating from {$^2A$} and in  panel (b) those originating
from {$^4A$}. The diagram {$^2A_2$} is obtained by differentiating with
respect to {${\tilde{\bf f} }(x')$} at position 2 {$or$} 3 in {$^2A$}. A
factor of 2 is gained. The diagram {$^2A_1$} comes from differentiating
with respct to {${\tilde{\bf f} }(x')$} at position 1. The diagram
{$^4A_4$} is obtained by differentiating with  respect to
{${\tilde{\bf f} }(x')$} at position 4 {$or$} 5 in {$^4A$}. A factor of 2
is gained. The diagrams {$^4A_3\ , ^4A_2$} and {$^4A_1$} come from
differentiating with respect to {${\tilde{\bf f} }(x')$} at positions 3, 2
and 1 respectively. Again the factor 1/2 remains at vertices that have
symmetry above them.
           }
             \label{20-fig4}
             \end{figure}

The calculation of the functional derivative using the diagrams in
Fig.~\ref{20-fig2} is straightforward. Every diagram having $n$ branches
of ${\bf u}_0$ contains a product of $n$ $\tilde{\bf f}\,$'s.  The
calculation of the derivative with respect to $\delta{\tilde{\bf f}} (x')$
means via the chain rule that we get $n$ contributions to $\delta{\bf
u}(x)/\delta{\tilde{\bf f}} (x')$.  Every such contribution is obtained by
dropping one of the branches of ${\bf u}_0$, and replacing it by branch of
${\bf G}^0$. An example of how this procedure is done for diagrams $^2A$
and $^4A$ in Fig.~\ref{20-fig2} is shown in Fig.~\ref{20-fig4}a and
Fig.~\ref{20-fig4}b respectively.

\begin{figure}
\hbox{
             \epsfxsize=3.1truein 
             \epsfbox{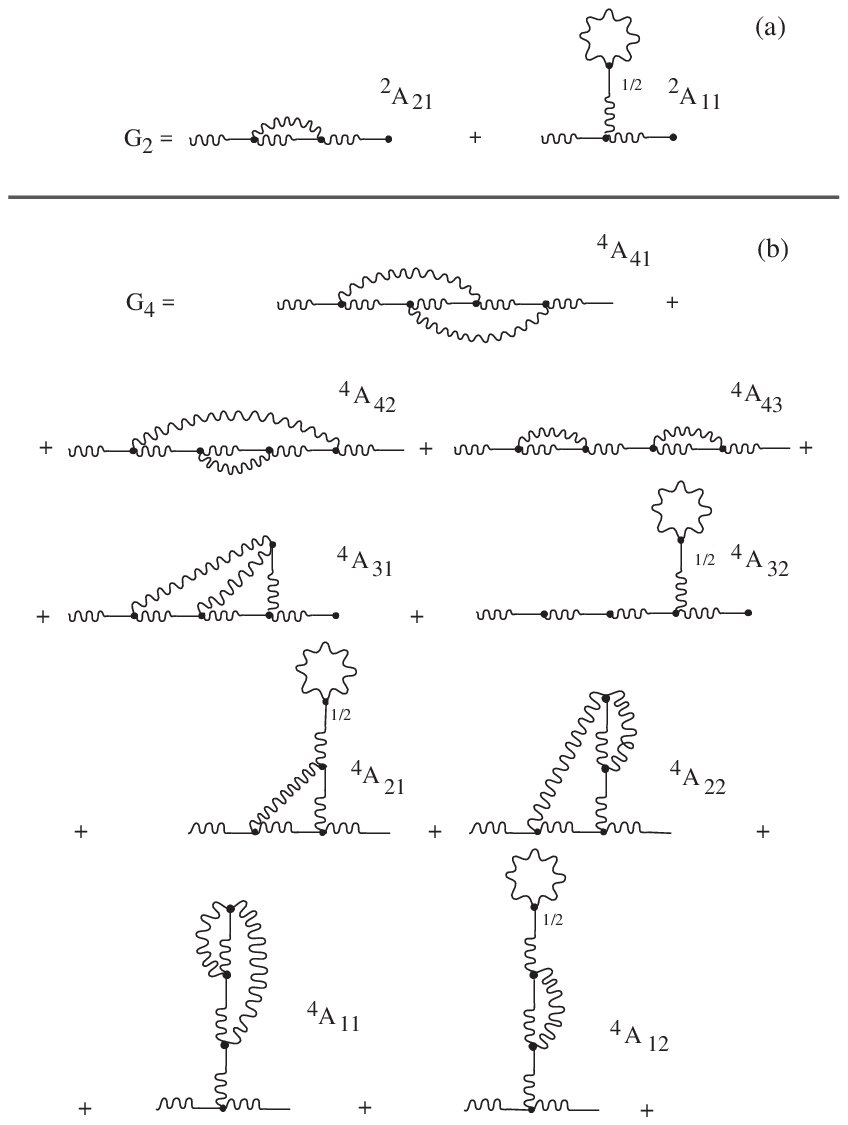}
                  \qquad 
                \epsfxsize=3.1truein 
             \epsfbox{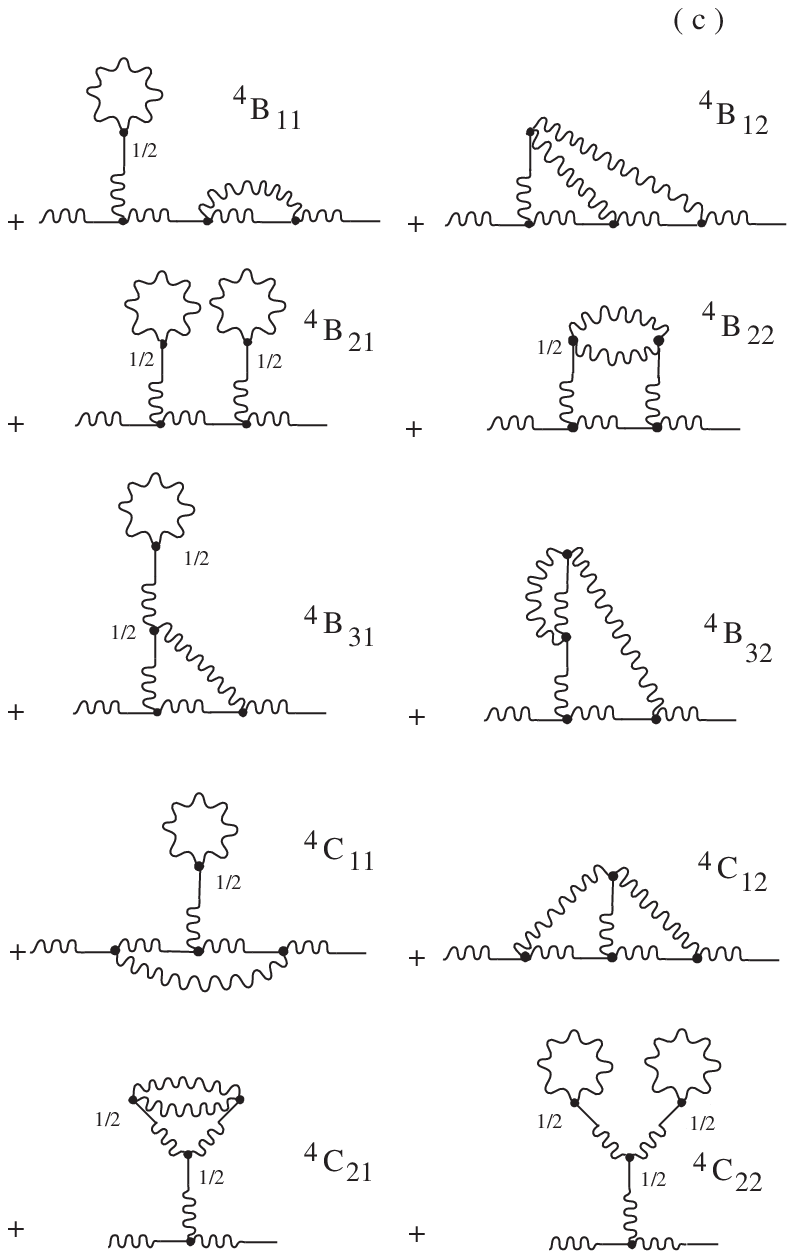}
} 
\caption{
Diagrams for the Green's function. Panel (a): The contributions to {${\bf
G}_2$}  originate from the trees shown in Fig.~4a by gluing two {${\bf
u}_0$} branches in the only possible way. Panel (b): The contributions to
{${\bf G}_4$} originating from the trees {$^4A_1-^4A_4$} shown in Fig.~4b.
The diagrams {$^4A_{41}-^4A_{43}$} come from {$^4A_4$} by gluing 1 - 3 and
2 - 4; 1 - 4 and 2 - 3 ; 1 - 2 and 3 - 4 repsectively.  The diagrams
 {$^4A_{31}$} come from {$^4A_3$} by gluing 1 - 4 and 2 - 5 {$or$} 1 - 5
and 2 - 4. A factor of 2 is gained.  {$^4A_{32}$} is obtained by gluing
{$^4A_3$} at positions 1 - 2 and 4 - 5.  The diagram {$^4A_{21}$} comes
from {$^4A_2$} by gluing 1-3 and 4-5.  {$^4A_{22}$} is obtained by gluing
1-4 and 3-5 {$or$} 1-5 and 3-4, gaining a factor of 2. Finally
{$^4A_{11}$} comes from {$^4A_1$} by gluing 2-4 and 3-5 {$or$} 2-5 and
3-4.  {$^4A_{12}$} obtains from {$^4A_1$} by gluing 2-3 and 4-5. (c)
Diagrams for {${\bf G}_4$} originating from {$^4B$} and {$^4C$} in
Fig.~2e.  Note that again all the numerical factors follow the same rules
as observed in Fig.~4.  }
\label{20-fig5}
\end{figure}

The second step is obtained by averaging the diagrams for $\delta{\bf
u}(x)/\delta{\tilde{\bf f}}(x')$ over realizations of ${\bf f} (x')$. We
can apply the rules (\ref{20d1}) to the averages of the diagram of
$\delta{\bf u}(x)/\delta{\tilde{\bf f}} (x')$ in a graphical sense. This
procedure is done in Fig.~\ref{20-fig5}. Fig.~\ref{20-fig5}a contains the
two contributions to ${\bf G}_2$, and Figs.~\ref{20-fig5}b and
\ref{20-fig5}c show all the 19 diagrams for ${\bf G}_4$. The diagram
denoted $^nA_{mp}$ are obtained from the diagrams $^nA_m$ in
Fig.~\ref{20-fig4}.  The diagrams denoted $^nB_{mp}$ and $^nC_{mp}$
originate from diagrams $^nB$ and $^nC$ in Fig.~\ref{20-fig2}.  Notice
that every diagram has an ``entry" which is the root of the tree in
Fig.~\ref{20-fig2}, and an ``exit" which is the branch ${\bf G}^0$ which
replaced a branch of ${\bf u}_0$. There is a unique path between entry and
exit which is composed of a chain of ${\bf G}^0\,$'s. We shall call this
path the ``principal path" of the diagram.  Notice that the entry always
begins with a wavy line, whereas the exit ends with a straight line. In
fact, all the diagrams appearing here can be drawn without reference to
the explicit derivation described here using simple topological rules.
However, since the diagrams described here are not in their final form we
defer the discussion of the appropriate rules for a later moment.

\subsubsection{The 2-point velocity correlation function}
The 2-point velocity correlation function ${\bf F}(x,x')$
is defined as
\begin{equation}
{F}_{\alpha\beta}(x,x')
=\left\langle u_\alpha(x) u_\beta(x')\right\rangle   \ .
\label{20d3}
\end{equation}
The calculation of this quantity is again based on the diagrammatic
expansion shown in Fig.~\ref{20-fig2}. To obtain ${\bf F}_{m+p}$ of order
$\Gamma^{m+p}$ we need first to take a contribution ${\bf u}_m$ and a
contribution ${\bf u}_p$ and mutiply them together. The $n$-th order ${\bf
F}_n$ is obtained as a sum of all ${\bf F}_{m+p}$ such that $m+p=n$.  In
the second step we average over the realizations of ${\tilde{\bf f}}$. As
before all odd order contributions vanish because they contain an
odd power of $\tilde{\bf f} $. For example,
\FL
\begin{eqnarray}
{\bf F}_0(x,x')&=&{\bf F}_{0+0}(x,x'),
\nonumber\\
{\bf F}_1(x,x')&=&{\bf F}_{1+0}(x,x')
+{\bf F}_{0+1}(x,x')=0\,,
\label{20d4}\\
{\bf F}_2(x,x')&=&{\bf F}_{2+0}(x,x')
+{\bf F}_{1+1}(x,x')+{\bf F}_{2+0}(x,x')\,,
\nonumber
\end{eqnarray}
where
\begin{equation}
 {\bf F}_{m+p}(x,x')=\left\langle{\bf u}_m(x){\bf u}_p(x')
\right\rangle\ .
\label{20d5}
\end{equation}

\begin{figure}
             \epsfxsize=13truecm
\centerline{             \epsfbox{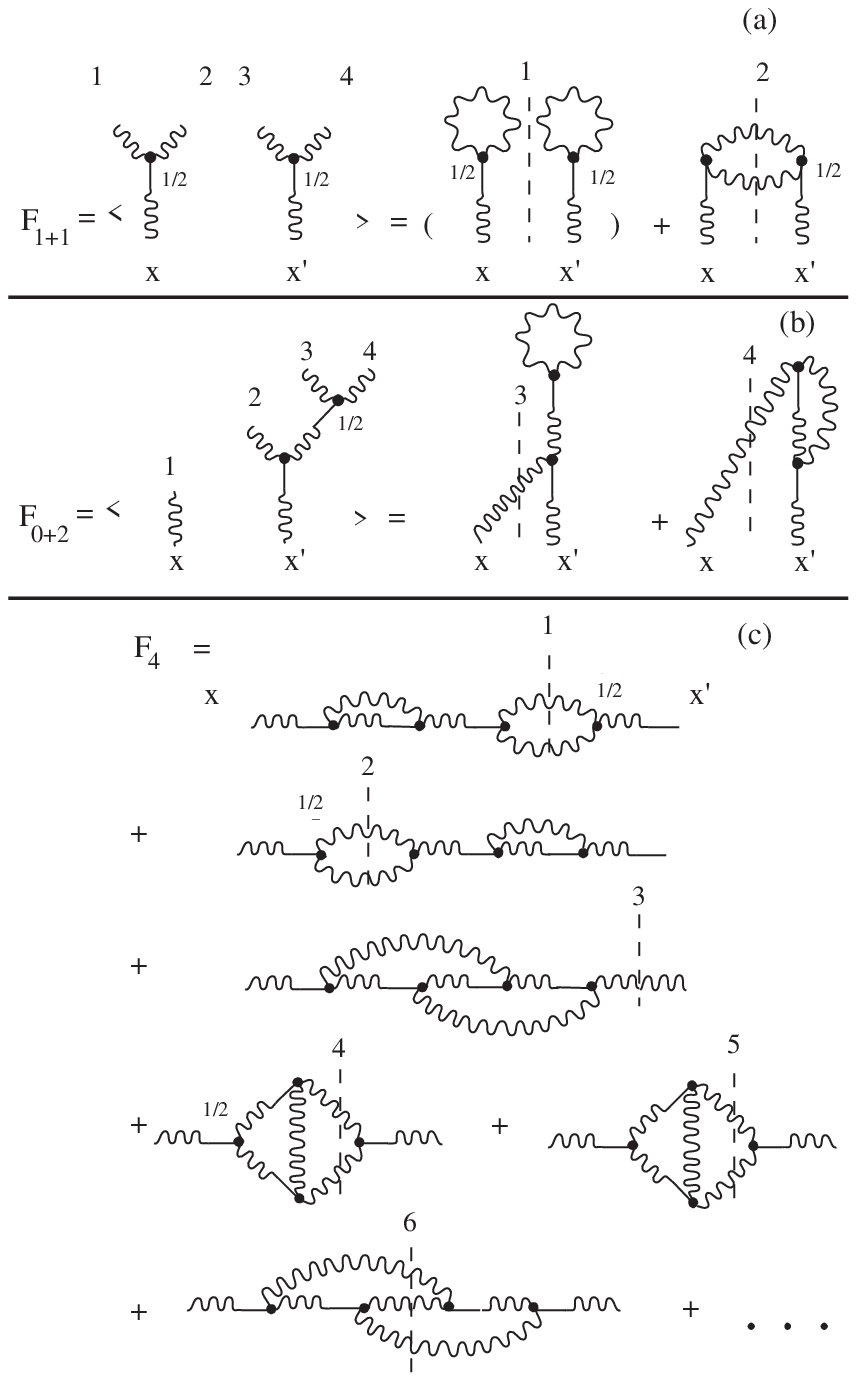} } 
             \vspace{.5cm}
\caption{
Diagrammatic representation of the 2-point velocity
correlation  function. Panel (a): {${\bf F}_{1+1}$} is obtained from the
the trees {${\bf u}_1(x)$} and {${\bf u}_1(x')$}.  Diagram 1 is obtained
by gluing 1-2 and 3-4, and diagram 2 by gluing 1-3 and 2-4 {$or$} 1-4
and 2-3, gaining a factor of 2. Panel (b): {$F_{0+2}$} is obtained from
the trees for {${\bf u}(x)$} and {${\bf u}_2(x')$}. Diagram 1 is obtained
by gluing 1-2 and 3-4, while the diagram 2 comes from gluing 1-3 and 2-4
{$or$} 1-4 and 2-3, gaining a factor of 2. Panel (c): Some typical
diagrams contributing to {${\bf F}_4$}.  Note that all the diagrams have a
``principal cross section" that separates trunks belonging to the left and
to the right trees. This cross section runs through 2-point correlators,
never through Green's functions. In contrast with the diagrams for the
Green's function that have wavy entries and straight exits, here we have
wavy entries on  both sides. The rules for the numerical factors of the
diagrams remain the  same.                   }
             \label{20-fig6}
             \end{figure}

To obtain the diagrammatic representation of this procedure we take one
tree for ${\bf u}_m(x)$ and one tree for ${\bf u}_p(x')$ from
Fig.~\ref{20-fig2}, and average the product of these trees according
to the rules (\ref{20d1}). In other words we need to pair the branches of
${\bf u}_0$ in all the possible ways, and to glue them as discussed
before in computing $\left\langle{\bf u}\right\rangle$ and the Green's
function.  The procedure is shown in Fig.~\ref{20-fig6}. Note that
every diagran has a uniquely defined "principal cross section" which
arises from the gluing of the two trees ${\bf u}_m(x)$ and ${\bf
u}_p(x)$.  We denote it in the diagrams as a vertical broken line, and we
draw the reader's attention to the fact that the principal cross section
cuts through correlators only, and not through Green's functions. This
fact is used later in the process of resummation.

This is the end of the naive perturbation theory. As we saw, it is an
expansion in powers of Re, and we are going to partly resum this expansion
to develop a reformulation with a better expansion parameter.  We
reiterate here that the procedure at this point seems very dependent on
the properties of the noise, since we used the rules (\ref{20d1}) time and
again. We defer further discussion of this issue until after the
resummations of various sorts, when we can take the limit ${\tilde{\bf f}}
\rightarrow 0$ with impunity.


\section{Resummations}
\label{20sect:resum}

In this chapter we discuss the resummation of the naive perturbation
theory that was developed in Chapter~\ref{20sect:naive}. We begin with the
mean velocity and its role in the resummation of the Green's function and
the correlator.

\subsection{The resummation of the mean velocity}
\label{20sect:resum-A}
In Section \ref{20sect:naive}B we discussed the diagrammatic series for
$\langle{\bf u}\rangle$ (see Fig.~\ref{20-fig3}) , and commented that it
has to sum up to zero when $\langle{\bf u}\rangle=0$.  In discussing the
diagrams for the Green's fucntion Fig.~\ref{20-fig5} and for the
correlator Fig.~\ref{20-fig6} we encounter again the same type of diagrams
that appear in $\langle{\bf u}\rangle$. For example in Fig.~\ref{20-fig5}
the diagrams $^2A_{11},~^4A_{32},~^4A_{21},~^4A_{11},~^4B_{21},~^4B_{31}$,
and $^4C_{11}$ all have a fragment which is the diagram for $\langle{\bf
u}_1\rangle$ in Fig.~\ref{20-fig3}a. In addition, the diagram $^4A_{11}$
has a fragment which is identical to $^3B_1$ in Fig.~\ref{20-fig3}. The
diagram $^4A_{12}$ has a fragment like $^3B_2$. The diagram $^4C_{21}$ has
a fragment like $^3A_2$, and lastly $^4C_{22}$ exhibits a fragment like
$^3A_1$.

All these diagrams that contain fragments belonging to $\langle{\bf
u}\rangle$ have a common feature. To see this denote the part of any
diagram that contains the principal path between entry and exit as the
``body"  of the diagram. Any fragment that can be disconnected from the
body by cutting one Green's function is called a ``weakly linked"
fragment. All the diagrams that we discussed in the previous paragraph
have weakly linked fragments. Consider now all the diagrams in
Fig.~\ref{20-fig5} that have two Green's functions in their principal
path. These are the diagrams $^2A_{11},~^4A_{11},~^4A_{12},~^4C_{21}$ and
$^4C_{22}$.  The sum of the weakly linked fragments of these diagrams is
exactly $\langle{\bf u}_1\rangle+\langle{\bf u}_3\rangle$ as can be seen
in Fig.~\ref{20-fig3}.  If we consider all the higher order diagrams for
${\bf G}$ which have two Green's functions in the principal path, we find
that their weakly connected fragments furnish all the remaining diagrams
in the series for $\langle{\bf u}\rangle$. The coefficients in front of
all these fragments is the same as the coefficient in the series for
$\langle{\bf u}\rangle$ since it is determined uniquely by the local
topology of the fragment, independently of the position that the fragment
occupies in the mother diagram. Accordingly all these diagrams with two
Green's functions in the principal path sum up to zero. The same story
repeats for all the diagrams that contain a weakly linked fragment. For
example the diagram $^4A_{21}$ in Fig~\ref{20-fig5}b is the first in the
series that exhibits a weakly linked fragment that eventually will be
resummed together with diagrams that have the same body, but higher order
weakly linked contributions that sum up to zero. The general conclusion is
that all the diagrams that have at least one weakly linked fragment sum
up to zero and need not be considered further in the resummed theory.

Next we discuss the appearance of $\langle{\bf u}\rangle$ in the series
for the 2-point correlation fucntion. In this series we find a new type
of diagram, like diagram 1 in Fig.~\ref{20-fig6}a. These are unlinked
diagrams which are obtained from averaging the left and the right tree
separately. Such diagrams contain no correlators that cross the principal
cross section.  Obviously such diagrams will resum to $\langle{\bf
u}\rangle^2$ which is zero. In addition we have diagrams with weakly
connected fragments. In the context of the diagrams for the 2-point
correlator we define the body of the diagram as the part that contains the
two entries (the roots of the original trees), which are denoted by ``$x$"
and ``$x'$'' in all the diagrams in Fig.~\ref{20-fig6}. A weakly linked
fragment is a fragment that can be disconnected from the body of the
diagram by cutting off one Green's function. An example of such a diagram
is diagram 1 in Fig.~\ref{20-fig6}b.  As in the case of the Green's
function, the infinite sets of weakly linked fragments with the same body
resums to $\langle{\bf u}\rangle=0$. From this point on we therefore
discard all the diagrams that have at least one weakly connected fragment.

\subsection{The Dyson resummation for the Green's function}
\label{20sect:resum-B}

In this section we discuss the Dyson line resummation of the series for
the Green's function. To this aim we classify the diagrams into three
classes.  The first class consists of diagrams with weakly linked
fragments which are all discarded. The other two classes are designated as
follows:

I. Principal path reducible diagrams. These are diagrams that can be
split into two disjoint pieces which contain more than one Green's
function, by cutting one bare Green's function that belongs to the
principal path. An example of such a diagram is $^4A_{43}$ in
Fig.~\ref{20-fig5}b. This diagram fall into two parts by cutting ${\bf
G}_0(x_2,x_3)$.

II. Principal path irreducible diagrams. These are the diagrams that
cannot be split as described in I. All the other diagrams in
Fig.~\ref{20-fig5} that do not have weakly linked fragments are principal
path irreducible.

\begin{figure}
             \epsfxsize=8.6truecm
\centerline{             \epsfbox{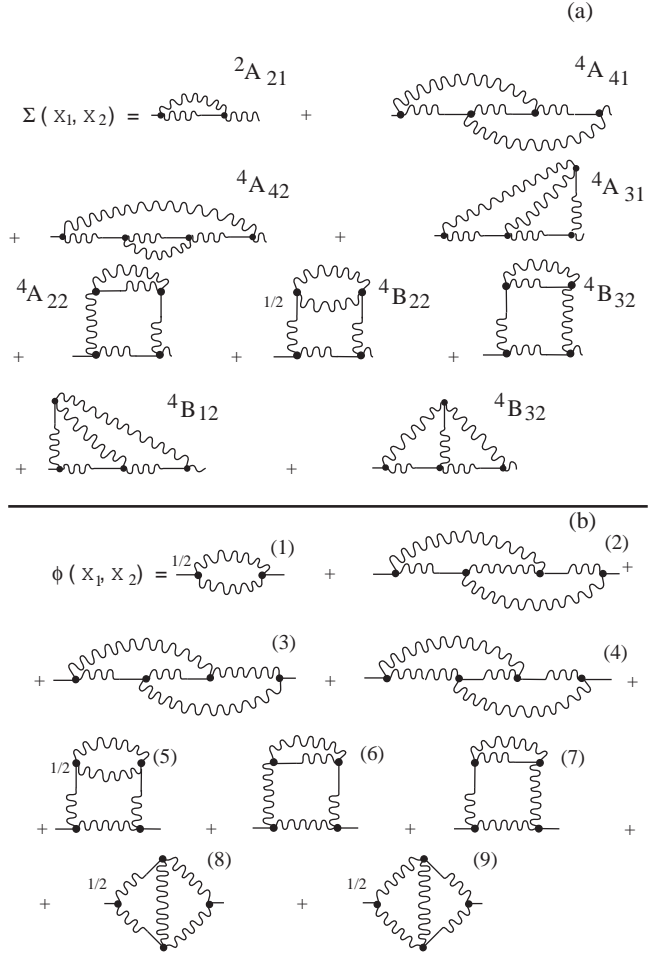} } 
             \vspace{.5cm}
\caption{
Panel (a):  Diagrammatic series for the {$\Sigma (x_1,x_2)$} operator.
All the  diagrams to order {$\Gamma ^4$} are shown. Every diagram is
denoted by the  same notation of the diagram it derives from in Fig.~5.
Panel (b): Diagrammatic series for the {$\Phi (x_1,x_2)$} operator. All
the diagrams to order $\Gamma ^4$ are shown. The numbering of the diagrams
is referred to in  the section on line resummation. }
\label{20-fig7}
\end{figure}

All the principal path irreducible diagrams, except ${\bf G}_0(x,x')$
itself, share the property that they start with a bare ${\bf G}_0(x,x_1)$,
they end up with a bare ${\bf G}_0(x_2,x')$, and in between they have a
principal path irreducible structure, say $S(x_1,x_2)$. The sum of all
these irreducible structures is defined as the ${\bbox{\Sigma}}$ operator,
which will be shown to contain all the information about the turbulent
eddy viscosity. Using the diagrams in Fig.~\ref{20-fig5} we develop the
diagrammatic expansion for ${\bbox{\Sigma}}(x,x_1)$ which is shown in
Fig.~\ref{20-fig7}. With the help of ${\bbox{\Sigma}}(x_1,x_2)$ we can say
that the sum of all the principal path irreducible diagrams is
\begin{eqnarray}
&&{\rm Sum~of~all~irreducibles}
\label{20e1} \\
&=&{\bf G}_0(x,x')+{\bf G}_0(x,x_1)*\Sigma
(x_1,x_2)\cdot {\bf G}_0(x_2,x')
\nonumber
\end{eqnarray}
where the star ``$*$"
and dot ``$\cdot$" products are as defined in the vertex  Eq.
(\ref{20b17}).  The operator ${\bbox{\Sigma}}(x_1,x_2)$ starts and ends
with a vertex, and it therefore connects with ``$*$" to the preceding
${\bf G}_0(x,x_1)$, and with a ``$\cdot$" to the following ${\bf
G}_0(x,x_1).$ Further summation of this series will be discussed soon.

The principal path reducible diagrams, of which we have only one
representative in Fig.~\ref{20-fig5}, can be also resummed using the
operator ${\bbox{\Sigma}}(x_1,x_2)$.
Note that the reducible diagram $^4A_{43}$ has
a structure such that that between $x_1$ and $x_2$ there exists the first
contribution to ${\bbox{\Sigma}} (x_1,x_2)$,
and after the point $x_2$ we see a
fragment that is identical to $^2A_{21}$, which is the first nonlinear
contribution to the Green's function.  The main observation, which is due
to Dyson, is that higher order reducible diagrams will have between $x_1$
and $x_2$ all the higher order terms in $\Sigma (x_1,x_2)$, and than after
$x_2$ we will have all the other nonlinear contributions to the Green's
function ${\bf G}(x_2,x')$. Therefore, we can write
\begin{eqnarray}
&&{\rm Sum~of~all~reducibles}
\label{20e2} \\
& = &{\bf G}_0(x,x_1)*\bbox{\Sigma}
(x_1,x_2)\cdot[ {\bf G}(x_2,x')-{\bf G}_0(x_2,x')]
\nonumber
\end{eqnarray}
Again we made use of the fact that all topologically possible diagrams
appear in the series, and that the numerical weight of each fragment is
only determined by its local symmetry, independent of its position in the
mother diagram.

Adding together (\ref{20e1}) and (\ref{20e2}) we get the Dyson equation
for the Green's function
\FL
\begin{equation}
{\bf G}(x,x') = {\bf G}_0(x,x')+{\bf
G}_0(x,x_1)*\Sigma (x_1,x_2)
\cdot{\bf G}(x_2,x')
\label{20e3}
\end{equation}
The formal solution of this equation in operator form is
\begin{equation}
\OG
=[ {\bf 1} -
\OG _0*
\OS ] ^{-1}
\OG_0
\label{20e4}
\end{equation}
or, more  explicitly
\begin{equation}
\OG= [
\OG_0^{-1}
- \OS] ^{-1}
=\Bigg[ {i\OP
\over   \partial/   \partial t-\nu{\nabla}^2
- \OP
\OS } \Bigg]
\label{20e5}
\end{equation}
where we insert the definition of $\OG_0$ from the Eq.  \ref{20b11}) has
been used. We see now that the $\OS$ operator serves as a renormalization
of the viscosity. The full appreciation of this fact will become clear in
subsection D.

\subsection{The Wyld resummation for the 2-point correlation function}
\label{20sect:resum-C}
\begin{figure}
             \epsfxsize=6truecm
\centerline{             \epsfbox{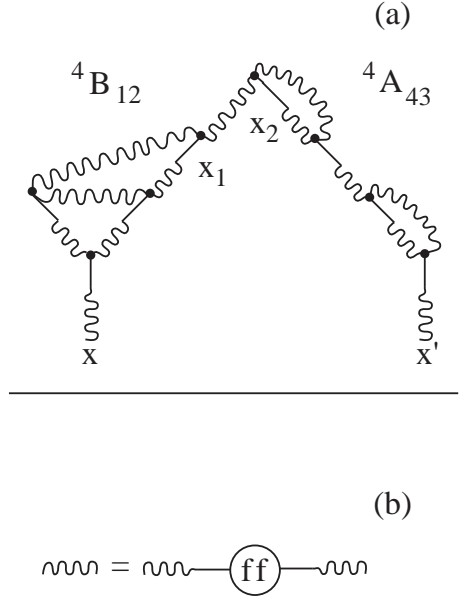} } 
             \vspace{.5cm}
\caption{
Panel (a). One of the diagrams that have one wavy line at the principal
cross section which appears in the series for the 2-point correlation
function. Panel (b). The representation of the lowest order contribution
to the  2-point correlator}
\label{20-fig8}
\end{figure}

In this section we discuss the Wyld line resummation of the series of the
2-point correlation function. We need again to classify the diagrams in
order to see clearly how to proceed. Firstly, all the diagrams that are
unlinked or have weakly linked fragments can be discarded. Next we
consider all the diagrams that have only one wavy line that crosses
the principal cross section. The diagram 2 in Fig.~\ref{20-fig6}b and the
diagram 3 in Fig.~\ref{20-fig6}c are such diagrams. In Fig.~\ref{20-fig8}
we show a typical 8$^{\rm th}$ order diagram of this type.

Recall now that the wavy line, which is a correlation function of two
${\bf u}_0$'s, $\langle{\bf u}_0(x){\bf u}_0(x')\rangle$, can be
interpreted as ${\bf G}_0(x,x_1)*\langle\tilde{{\bf f} }(x_1)\tilde{{\bf
f} }(x_2)\rangle *{\bf G}_0(x_2,x')$ (see Fig.~\ref{20-fig8}b). Thus, on
the two sides of the $\langle\tilde{{\bf f} }(x_1)\tilde{{\bf f}
}(x_2)\rangle$ we have two typical contributions to the diagrammatic
series of ${\bf G}$ that was discussed in the last section.  On both sides
of the principal cross section we get, excluding $\langle\tilde{{\bf f}
}(x_1)\tilde{{\bf f} }(x_2)\rangle$, portions of diagrams that have an
entry and an exit like all the diagrams in Fig.~\ref{20-fig5}. Thus the
diagrams 2 in Fig.~\ref{20-fig6}b and  3 in Fig.~\ref{20-fig6}c have
fragments which are identical to $^2A_{21}$ in Fig.~\ref{20-fig5}a. In the
example of Fig.~\ref{20-fig8} the two fragments are identical with
diagrams $^4B_{12}$ and $^4A_{43}$ in Fig.~\ref{20-fig5}.  All the other
diagrams with only one wavy line across the principal cross sections will
have all the other contributions to the series of ${\bf G}$ on both sides
of the $\langle\tilde{{\bf f} }(x_1)\tilde{{\bf f} }(x_2)\rangle$
correlation. Thus we conclude that

\begin{eqnarray}
&&{\rm
Sum~of~all~diagrams~with~one} \nonumber \\ &&{\rm
wavy~line~at~the~cross~section} \nonumber \\ &=& {\bf
G}(x,x_1)*\langle{\tilde{\bf f} }(x_1) {\tilde{\bf f}} (x_2)\rangle *{\bf
G}(x_2,x')\ .
\label{20f1}
\end{eqnarray}
Next we consider the diagrams with two or more wavy lines crossing the
principal cross section. Examples are diagram 2 in Fig.~\ref{20-fig6}a,
and diagrams 1,2,4, 5 and 6 in Fig.~\ref{20-fig6}c. Of these diagrams
consider first all those that cannot be split into two parts (containing
more than a single Green's function) by cutting off one Green's function.
All these diagrams have two entries via straight lines on the right and
the left, and a structure, say $\phi (x_1,x_2)$ in between. The sum of all
these $\phi (x_1,x_2)$ contributions is defined as the $\Phi (x_1,x_2)$
operator. In Fig.~\ref{20-fig7}b all the contributions of $\Phi (x_1,x_2)$
up to 4$^{\rm th}$ order are exhibited.

All the rest of the diagrams that have two or more wavy lines crossing the
principal cross section are analyzed as follows: start from the principal
cross section, and move left until you meet for the first time a Green's
function that can be cut such that the diagram splits into two disjoint
parts. Denote the coordinate of the vertex at the exit of this Green's
function as $x_1$. Start again from the the principal cross section and
move the right until you meet another Green's function that can be cut
such that the diagram splits into two parts. Denote the coordinate at the
vertex at the exit by $x_2$. This notation is shown in all the relevant
diagrams of Fig.~\ref{20-fig6}. The fragment to the left of $x_1$ and to
the right of $x_2$ is a typical fragment that appears in the series for
${\bf G}$.  For example in diagram 2 Fig.~\ref{20-fig6}a we have a ${\bf
G}_0(x,x_1)$ on the left of $x_1$ and ${\bf G}_0(x_2,x')$ on the right of
$x_2$. In diagram 1 in Fig.~\ref{20-fig6}c we have a fragment $^2A_{21}$
(Fig.~\ref{20-fig5}) to the left of $x_1$ and again ${\bf G}_0$ to right
of $x_2$.  In the diagram 2 in Fig.~\ref{20-fig6}c we have the inverted
situation. Obviously, in the higher order diagrams we will find all the
remaining contributions to ${\bf G}$. We thus conclude that all these
diagrams which have two or more wavy lines crossing the principal crosss
section may be resummed to
\begin{eqnarray}
&&{\rm Sum~of~all~diagrams~with~2~and~more }
\nonumber\\
&&{\rm wavy~lines~ crossing~ the~ cross~ section }
\label{20f2} \\
& =&{\bf G}(x,x_1)*\Phi (x_1,x_2) *{\bf G}(x_2,x') \ .
\nonumber
\end{eqnarray}
Summing up Eqs.(\ref{20f1}) and (\ref{20f2}) we recover the Wyld equation
\begin{eqnarray}
{\bf F}(&x,x'&)
\label{20f3} \\
&=&
{\bf G}(x,x_1)*[ \langle{\tilde{\bf f} }   (x_1)
{\tilde{\bf f} }(x_2)\rangle + {\bbox{\Phi}}(x_1,x_2)]
*{\bf G}(x_2,x')\ .
\nonumber
\end{eqnarray}
This equation, after the resummation of the series for ${\bbox{\Phi}}$,
will allow us to discuss the limit ${\tilde{\bf f} }\rightarrow 0$.
\subsection{Line resummation}
\label{20sect:resum-D}
Up to now our perturbation series is in orders of the Re number, as
explained in Sect.\ref{20sect:naive}. The Green's function and the
2-point correlators which appear in the diagrams for ${\bbox{\Phi}}$ and
${\bbox{\Sigma}}$ are the bare ones. In this section we perform the so
called ``line resummation" which results in diagrams in which all the
propagators are dressed ones, and at the same time the perturbation
parameter becomes of the order of 1.

To accomplish this resummation we introduce a few notions.  First is the
notion of the ``end points" of a diagram. The end points are defined as
the entry and exit in the case of the dagrams for ${\bbox{\Sigma}}$,
and the two exits in the case of diagrams for ${\bbox{\Phi}}$. Next we
define ``one eddy reducible" diagrams and ``one-eddy irreducible"
diagrams. These types of diagrams play a topologically similar role to the
``one-particle reducible and irreducible" diagrams in the standard Feynman
diagrams.  The definition is as follows: Name as a ``one-eddy reducible"
fragment any fragment that can be completely disjoined from the part of a
diagram that contains the end points by cutting any two lines, be whether
Green's fucntions or correlators. The part that contains the two end
points will be called the bulk of the diagram, which may or may not be the
same as the body of the diagram as defined in Section\ref{20sect:resum-A}.
The convention will be that the fragment contains the two lines that have
been cut. In other words, we cut lines near the vertex that
belongs to the rest of the diagram. It follows from this convention that
one-eddy reducible diagrams are at least of order $\Gamma^2$. Define now

I. One-eddy reducible diagrams: diagrams that contain at least one
one-eddy reducible fragment.

II. One-eddy irreducible diagrams: diagrams that contain no
one-eddy reducible fragment.

Examples for one-eddy reducible diagrams are diagrams $^4A_{42}$,
$^4A_{22}$, $^4B_{22}$ and $^4B_{32}$ in Fig.~\ref{20-fig7}a for
$\bbox{\Sigma}$, and diagrams 5, 6 and 7 in Fig.~\ref{20-fig7}b for
$\bbox{\Phi}$. The rest of the diagrams in Fig.~\ref{20-fig7} are one-eddy
irreducible.

We can classify one-eddy reducible fragments into two classes, called as
class F and class G. Class F has two wavy ends (like the fragment of
$^4A_{22}$) and class G has one wavy and one straight end (like the
fragment of $^4A_{42}$).  An observation of principal significance is that
the sum of all the possible fragments of class F is precisely the
nonlinear correction to the 2-point correlator F, whereas the sum of all
the possible fragments of class G is the nonlinear correction to the
Green's function G. As an example the fragment of $^2 A_{42}$ is the first
nonlinear correction to ${\bf G}$ which is $^2A_{21}$ in
Fig.~\ref{20-fig5}a. The fragments of $^4B_{22}$ in the series for
${\bbox{\Sigma}}$, Fig.~\ref{20-fig7}a and the fragment of diagram 5 in
the series for ${\bbox{\Phi}}$, Fig.~\ref{20-fig7}b are the same, and are
the nonlinear correction 2  for ${\bf F}$ in Fig.~\ref{20-fig6}a.  The
fragment of the diagram $^4A_{22}$ is the diagram 2 in
Fig.~\ref{20-fig6}b.

Consider now the irreducible diagram $^2A_{21}$ in Fig.~\ref{20-fig7}a. By
summing it together with the one-eddy reducible diagrams whose bulk part
contains either one bare correlator or one bare Green's function we obtain
the diagram 1 in Fig.~\ref{20-fig9}a. Similarly, every irreducible diagram
in the series for ${\bbox{\Sigma}}$ and ${\bbox{\Phi}}$ acts as the
beginning of an infinite series of one-eddy reducible diagrams that
renormalize all bare propagators to dressed ones. In Fig.~\ref{20-fig9} we
show all the contributions up to 4'th order of the resulting series. This
(very important) step in the theory is available due to two facts: the
first is that all the topologically allowed diagrams are present,
and the second is that the weight of each fragment is determined locally
by its symmetry properties.

We note that at this point one can state the following rules of how to
draw any diagram in the renormalized series for ${\bbox{\Sigma}}$ and
${\bbox{\Phi}}$:

I. Only diagrams with an even number of vertices appear.

II. The series for ${\bbox{\Sigma}}$ and ${\bbox{\Phi}}$ contains all
the topologically allowed diagrams made of $2n$ vertices with one straight
and two wavy legs.  The difference is that the diagrams for
${\bbox{\Sigma}}$ must have an entry with wavy line and an exit with a
straight line, whereas diagrams for ${\bbox{\Phi}}$ have two exits with
straight lines.

III. One needs to exclude all the one-eddy reducible diagrams, and
also all diagrams that include closed loops of Green's functions. The
latter are zero by causality. They never appear in the method of
derivation discussed above, but  they do appear in the path integral
formulation that is discussed in Section \ref{20sect:pathint}. They
vanish also there, of course. From these properties it follows that the
diagrams for ${\bbox{\Sigma}}$ have a unique principal path, and that the
diagrams for ${\bbox{\Phi}}$ have a principal cross section.

IV. The coefficient in front of a diagram is determined by the symmetry
properties. Every flip of legs that leaves the diagram invariant
contributes a factor of 1/2. However, the one-eddy irreducible diagrams
for ${\bbox{\Sigma}}$ cannot have any element of symmetry, and therefore
all their weights are unity. The irreducible diagrams for F can have at
most one element of symmetry, and they will have a weight of 1/2. Examples
are diagrams 1,5 and 6 in Fig.~\ref{20-fig9}b.

We can show now that the effective coupling constant is of the order of
unity. First we discuss how many correlators, propagators and vertices
appear in a $2n$'th order diagram for ${\bbox{\Sigma}}$ and
${\bbox{\Phi}}$. In a symbolic fashion we can write that
\begin{eqnarray}
\Sigma_{2n}&\sim&\Sigma_2\Lambda^{n-1}\,,
\quad \Sigma_2 \sim(\Gamma^2GF)\ ;
\label{20g1}\\
\Phi_{2n}&\sim&\Phi_2\Lambda^{n-1}\,,\quad
\Phi_2\sim\Gamma^2F^2
\label{20g2}
\end{eqnarray}
where the new expansion parameter $\Lambda$ is of the order of
\begin{equation}
\Lambda\sim\Gamma^2G^2F\ .
\label{20g3}
\end{equation}
To estimate the order of magnitude of $\Lambda$ we notice that $G$,
according to the Dyson equation (\ref{20e4}), is of the order of
$\Sigma^{-1}$.  Using this in (\ref{20g3}) we get
\begin{equation}
\Lambda\sim{\Gamma^2GF\over\Sigma}
\sim{\Gamma^2GF\over\Gamma^2GF}\sim 1\ .
\label{20g4}
\end{equation}
where Eq. (\ref{20g1}) has been used. We see that the line-renormalization
has shifted our expansion parameter from the order of Re to the order of
1.

At this point one usually considers the possibility of vertex
renormalization.  In our context this can be achieved by looking at the
2-eddy reducible fragments that can be cut from the bulk by severing three
legs. It will turn out that our perturbation series has the deeply
non-trivial property that that this step is not needed in the final
presentation of the theory.  This important issue is discussed further in
\cite{93LL}.

We thus end with the Wyld equation which can be rewritten as
\begin{eqnarray}
{\bf F}(x,x') &=&{\bf G}(x,x_1)*
\Big[ \langle{\bf f} (x_1){\bf  f} (x_2)\rangle
\label{20g5}\\
&&+\langle{\bbox{\phi}} (x_1) {\bbox{\phi}} (x_2)\rangle +{\bbox{\Phi}}
(x_1,x_2)\Big] *{\bf G}(x_2,x')
\nonumber
\end{eqnarray}
where we have made use of the fact that the thermal noise ${\bf f} $ is
uncorrelated with the stirring force $\bbox{\phi}$. We are going to seek
a solution of (\ref{20g5}) for which the energy of the turbulent motion
per mode is much larger than the order of $k_{_{\rm B}}T$.  For such
solutions the effect of the thermal noise on the observed statisitcs of
the velocity fluctuations will be negligible, and we can omit
$\langle{\bf f} (x_1){\bf f} (x_2)\rangle$ from (\ref{20g5}) with
impunity.  On the other hand, the role of the stirring force $\langle
\bbox{\phi} (x_1)\bbox{\phi} (x_2)\rangle$ can be very important. In
general the observed statistics depends on how the fluid is stirred. The
choice of stirring that models the boundary condition of experimental
turbuelence is discussed in \cite{95LP-b} after performing some of the
analysis on the operators ${\bbox{\Sigma}}$ and ${\bbox{\Phi}}$.

\subsection{Intuitive meaning of the Dyson and Wyld equations}
\label{20sect:resum-E}

\begin{figure}
             \epsfxsize=8.6truecm
\centerline{             \epsfbox{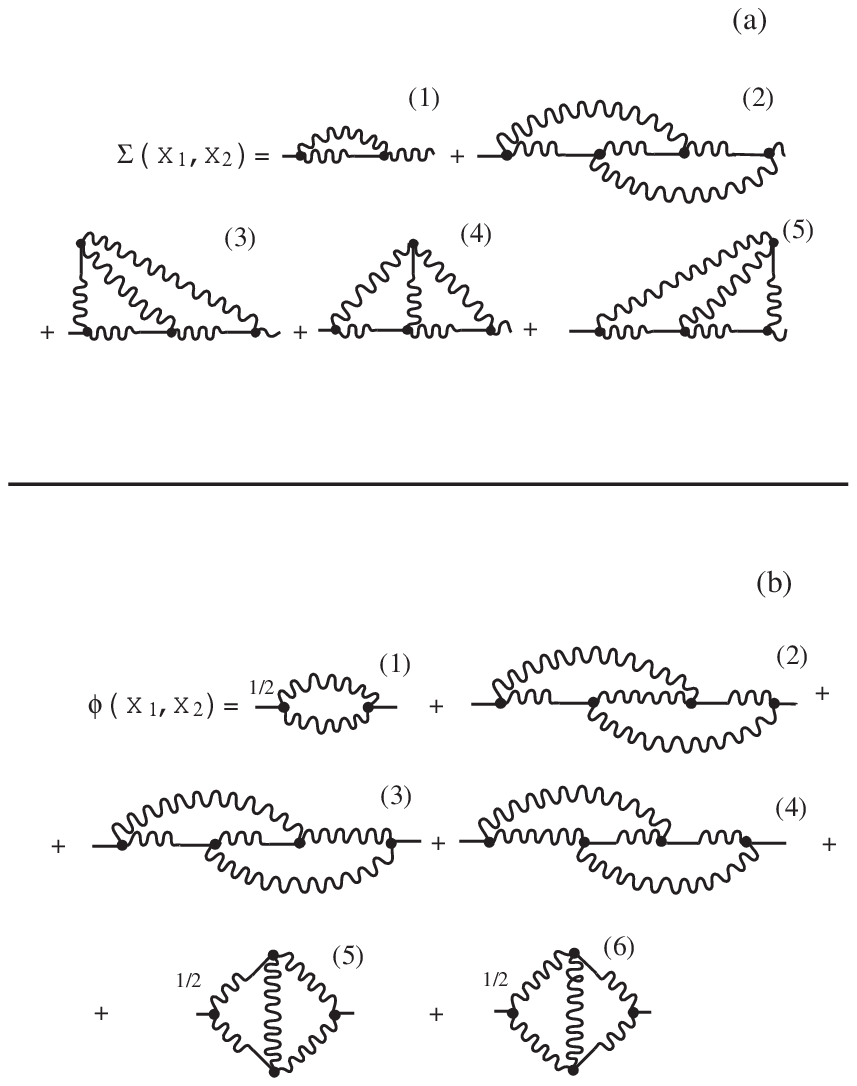} } 
             \vspace{.5cm}
\caption{
Diagramatic representation of of the operators {$\Sigma$} and
{$\Phi$} in the  line-renormalized theory. The symbols for the
Green's function and  the 2-point correlators are bold; they are
renormalized to all orders.      }
             \label{20-fig9}
             \end{figure}

The physical significance of the Dyson equation (\ref{20e3}) is that the
dressed response, which is a function of ${\bf r},{\bf r}'$ and $t-t'$, is
determined by hydrodynamic interactions involving intermediate points.
For example, the response to forcing at ${\bf r}'$ has one direct
contribution at $\bf r $, which is ${\bf G}_0(x,x')$. However, for any
finite time the response to forcing at ${\bf r}'$ is mediated by
interactions at points ${\rm r}_1$ which via ${\bbox{\Sigma}}$ appear at
point ${\bf r}$.  In the one loop approximation, which is diagram 1 in
Fig.~\ref{20-fig9}, ${\bbox{\Sigma}}$ itself has a Green's function that
mediates directly between ${\bf r}_1$ and ${\bf r}$. In higher order
contributions to ${\bbox{\Sigma}}$ there are sequential contributions due
to forcing at ${\bf r}'$ that are mediated by responses at ${\bf
r}_1,{\bf r}_2$ ...  until ${\bf r}$ is reached.  ${\bbox{\Sigma}}$
represents the dressed response which is the sum of all these sequential
responses at multiple intermediate sets of points. The Green's functions
that mediate intermediate points are weighted by the correlators of
velocity differences between these points; if these correlators are small,
the contribution of the Green's fucntion to the total response is also
small.

The intuitive understanding of the Wyld equation is also straightforward.
{}From the equation of motion (\ref{20b10}), written schematically as
\begin{equation}
{\bf u}={1\over[    \partial/   \partial t-\nu{\nabla}^2]}
\{ -[ {\bf u}\cdot  {\bbox\nabla}] {\bf u} +{\tilde{\bf f} }\}\,,
\label{20h1}
\end{equation}
we see that the nonlinear term $[ {\bf u}\cdot{\bbox\nabla}]\,{\bf u}$
can be understood as an ``additional noise" in the equation of motion.
It is natural to expect that the double correlator of the forcing will
have a nonlinear contribution of the type
\begin{eqnarray}
&&\langle
{\tilde{\bf f} }({\rm r}_1,t_1){\tilde{\bf f}}(r_2,t_2) \rangle_{_{\rm
NL}} \label{20h2}\\ &\sim &\langle[{\bf u}({\rm
r}_1,t_1)\cdot{\bbox{\nabla}}_1] {\bf u}({\rm r}_1,t_1)[ {\bf
u}(r_2,t_2)\cdot {\bbox\nabla}_2{\bf u}] ({\rm   r}_2,t_2)  \rangle\ .
\nonumber
\end{eqnarray}
Indeed, the mass operator $\Phi_{\alpha\beta}({\bf r},{\bf r}',t-t')$ can
be written exactly  as
\FL
\begin{eqnarray}
&&\Phi_{\alpha\beta}({\bf r},{\bf r}',t-t')
\label{20h3} \\
&=&{ \partial\over   \partial r_\gamma} {   \partial\over
\partial r_\delta^{\prime}}F_{\alpha\gamma\beta\delta}({\bf r},{\bf r},
{\bf r}',{\bf r}',t,t,t',t')\,,
\nonumber
\end{eqnarray}
where $F_{\alpha\gamma\beta\delta}$ is the fourth order correlation of
${\bf u}$.  Thus Eq.(\ref{20f3}) can be understood as the result of
squaring Eq.(\ref{20h1}) and averaging, up to the dressing of the Green's
functions. The content of the Wyld diagrammatics is the dressing of the
bare Green's function that appears in Eq.(\ref{20h1}), and the
representation of the fourth order $F_{\alpha\gamma\beta\delta}$ in terms
of second order $F_{\alpha\beta}$. The dressing of the Green's function
appears from the cross terms between $[ {\bf u}\cdot{\bbox\nabla}] {\bf
u}$ and $\tilde{{\bf f} }$ in (\ref{20h1}). Note that the analytic form of
the $\Phi_{\alpha\beta}$ in the 1-loop order, diagram 1 in
Fig.~\ref{20-fig9}b, follows directly from the Gaussian deomposition of
$F_{\alpha\gamma \beta\delta}$.

\begin{figure}
             \epsfxsize=8.6truecm
\centerline{            \epsfbox{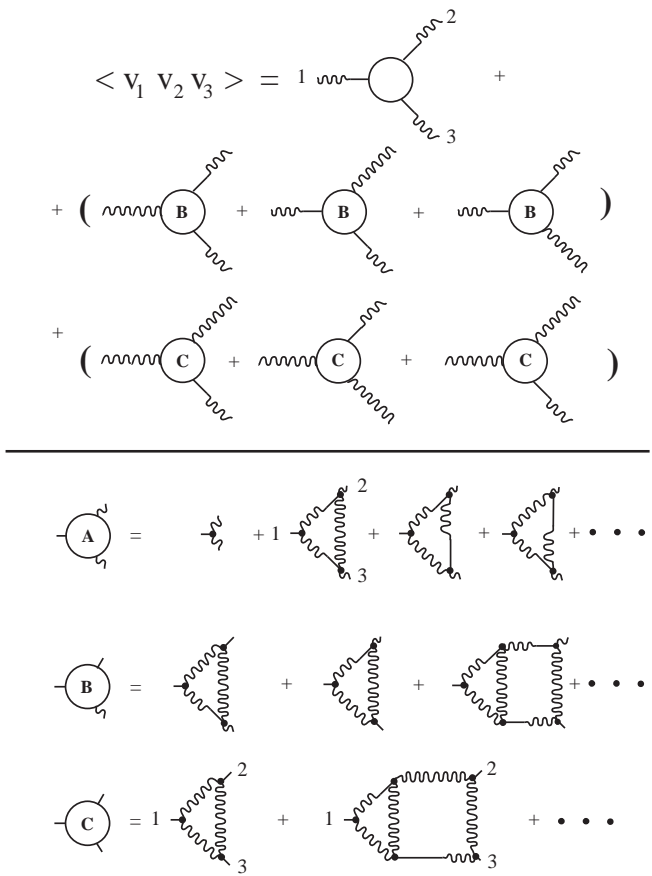} } 
             \vspace{.5cm}
\caption{  Diagrammatic representation of the 3-point velocity
correlator. Panel (a): There are 3 types of dressed vertices. Panel (b):
The diagrammatic  expansion for the three vertices appearing in (a). The
expansion is in  terms of dressed propagators and bare vertices, and we
show all the  diagrams up to fifth order in the bare vertex. }
             \label{20-fig10}
             \end{figure}

\subsection{3-point and higer order velocity correlation functions}
\label{20sect:resum-F}

To calculate the $n$-point velocity correlation function we need to
take $n$ trees from Fig.~\ref{20-fig2} and average their product.
Diagrammatically it means that we pair branches in all the possible ways,
and glue. Next we need to resum all the unlinked diagrams (that sum up to
zero), and all the weakly linked diagrams (with the body defined as the
structure having n entries) which also give zero contribution. Next we
perform a line-renormalization. The procedure is identical to the one
described above, and we leave the details to the interested reader. The
final result for the 3-point correlation function is shown in
Fig.~\ref{20-fig10}. There appear three types of vertices that we denote
as A,B and C. Their diagrammatic series is shown in Fig.~\ref{20-fig10}b.
The series in 10b is in terms of bare vertices, since, as we said before
there is no need to renormalize the vertices in this theory (see
\cite{93LL}).
\begin{figure}
             \epsfxsize=8.6truecm
\centerline{             \epsfbox{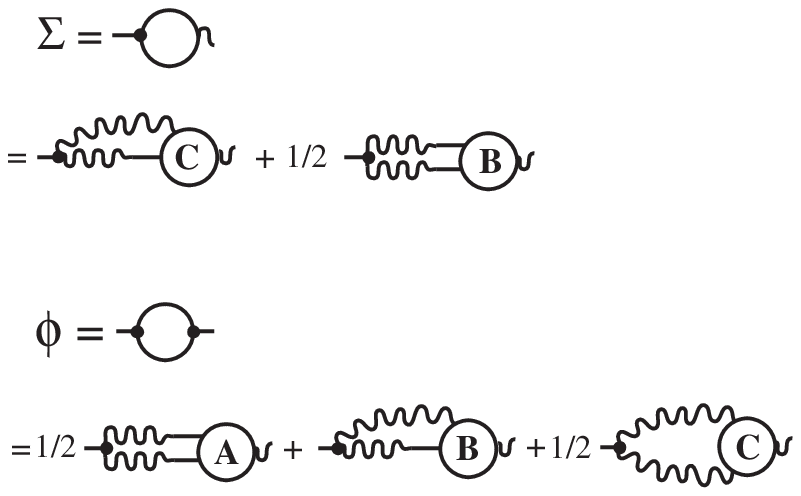} } 
             \vspace{.5cm}
\caption{
The condensed representation of the operators {$\Sigma$} and
{$\Phi$} in terms of the dressed vertices defined in Fig.~10b. }
             \label{20-fig11}
             \end{figure}

It should be noted that the objects A, B and C can be used to condense the
series for the operators ${\bbox{\Sigma}}$ and ${\bbox{\Phi}}$. This is
shown in Fig.~\ref{20-fig11}.  The formal elegance of this condensed
presentation is of no known use in the analysis.

\section{Functional Integral Formulation}
\label{20sect:pathint}
\subsection{Introduction}
\label{20sect:pathint-A}
In this chapter we discuss the path integral formulation of the
statistical theory of Navier-Stokes turbulence. The path integral
formulation does not lead to different diagrammatics than the one
described and discussed in the last chapter. Its main advantage is
two-fold. On one hand it allows a very compact representation, and
on the other hand it offers the ability to effect an infinite partial
resummation by changing variables in the path integral. This latter
property is the one that will turn out to be very useful for us, allowing
us to explore new renormalizations of the diagrammatic series which are
different from the Wyld and Dyson resummations, see \cite{93LL}

\subsection{Setting up the generating functional}
\label{20sect:pathint-B}
The starting point for the statistical description of turbulence in
the functional approach is the same equation of motion used in the
Wyld direct diagrammatic expansion:
\begin{equation}
\partial{\bf u}/   \partial t+({\bf u}
\cdot{\bbox\nabla} ){\bf u}-\nu{\bbox\nabla} ^2{\bf u}-{\bbox\nabla}
p={\tilde{\bf f} }\,,
\label{20i1}
\end{equation}
with the same properties of the random force ${\tilde{\bf f} }$. As
before, we need to find the statistical quantities by averaging over all
the realizations of the random force ${\tilde{\bf f} }$.

Introduce the Navier Stokes functional of ${\bf u}$, ${\cal N}
\lbrace{\bf u}(x)\rbrace$ as
\begin{equation}
{\cal N}\lbrace{\bf u}(x)\rbrace\equiv   \partial{\bf u}/   \partial
t+({\bf u}\cdot{\bbox\nabla}){\bf u}
-\nu{\nabla}^2{\bf u}-{\bbox\nabla}  p\ .
\label{20i2}
\end{equation}
Given any functional of ${\bf u}$, say ${\cal M}\lbrace{\bf u}(x)\rbrace$, we
can calculate its value for a given realization of ${\bf u}$ which in turn
is determined by some realization of the random force $\tilde{{\bf f} }$ .
This is done using the formal solution of Eqs.~(\ref{20i1})--(\ref{20i2})
\begin{equation}
{\bf u}(x)={\cal N}^{-1}\lbrace{\tilde{\bf f} }(x)\rbrace
\label{20i3}
\end{equation}
according to
\begin{equation}
{\cal M}\lbrace{\bf u}(x)\rbrace ={\cal M}\big\lbrace{\cal N}^{-1}\lbrace
{\tilde {\bf f} }
(x)\rbrace\big\rbrace\ .
\label{20i4}
\end{equation}
The RHS of this equation can be rewritten in the equivalent form
\begin{eqnarray}
&&{\cal M}\big\lbrace{\cal N}^{-1}\lbrace\tilde{{\bf f}
}(x)\rbrace\big\rbrace
\label{20i5} \\
&=&\int{\cal M}\lbrace{\bf u}(x)\rbrace\delta \Big[{\bf
u}(x)-{\cal N}^{-1}\{ \tilde{\bf f}(x)\} \Big]
D{\bf  u}(x)\ .
\nonumber
\end{eqnarray}
Here the functional integration is interpreted as the continuum limit
\begin{equation}
\int D{\bf u}(x)=\lim_{m\rightarrow\infty}\prod^M_{i=1}\int
d^3u(x_i)\delta \big[{\bbox\nabla}\cdot{\bf
u}(x_i)\big]\,,
\label{20i6}
\end{equation}
where $\lbrace x_i\rbrace =\lbrace{\rm r}_i,t_i\rbrace$ and $i$ runs over
a discrete space-time grid and  $M$ is the total number of points in the
4-dimensional grid.  The delta function
$\delta [{\bbox\nabla}\cdot{\bf u}(x_i)]$ limits the integration to
divergenceless contributions. The delta functional in (\ref{20i5}) is
defined as the continuum limit
\begin{eqnarray}
&& \delta \Big[{\bf u}(x)-{\cal N}^{-1}\lbrace{\tilde{\bf f} }(x)
\rbrace \Big]
\label{20i7} \\
&=&\lim_{M\rightarrow\infty}\prod^M_{i=1}
\int\delta \Big[{\bf u}(x_i)-{\cal N}^{-1}\lbrace{\tilde{\bf f} }(x_i)
\rbrace   \Big]\ .
\nonumber
\end{eqnarray}
 Note that Eq.~(\ref{20i5}) is but a cumbersome generalization to the
continuum of  the obvious equation
\begin{equation}
f(x)=\int f(y)\delta (x-y)dy\ .
\label{20i8}
\end{equation}
Remembering that
\begin{equation}
\delta \big[ g^{-1}(y)-x\big]=(dg/dx)\delta\big[  g(x)-y\big]\,,
\label{20i9}
\end{equation}
one may generalize to the continuum  limit according to
\FL
\begin{equation}
\delta \big[{\bf u}(x)-{\cal  N}^{-1}
\lbrace{\tilde{\bf f} }(x)\rbrace\big]
=J\lbrace{\bf u}\rbrace\delta \big[{\cal N} \lbrace{\bf u}(x)\rbrace
-{\tilde{ \bf f}}(x)\big],
\label{20i10} \end{equation}
where the Jacobian $J\lbrace{\bf u}\rbrace$
is given by
\begin{equation} J\lbrace{\bf u}\rbrace ={\rm det}
\Bigg|{\delta{\cal N}_\alpha\lbrace{\bf u}(x)\rbrace\over\delta u_\beta
(y)}\Bigg|\ .  \label{20i11}
\end{equation}
On the grid this Jacobian is the determinant of the $3M\times 3M$ matrix
\begin{eqnarray}
&&{\delta{\cal N}_\alpha \lbrace {\bf
u}(x_i)\rbrace\over\delta u_\beta (x_j)}
\label{20i12} \\
&=&\Big\{
\delta_{\alpha\beta} \Big[ {\partial \over   \partial t}+({\bf
u}\cdot{\bbox\nabla} ) -\nu{\nabla}^2\Big] + \partial_\beta u_\alpha
\Big\} \delta (x_i-x_j)\ .
\nonumber
\end{eqnarray}
In the Appendix we demonstrate that this determinant is  unity.  Thus
Eq.(\ref{20i4}) for ${\cal M}$ at a given realization of ${\tilde{\bf f}
}$  can be rewritten as
\FL
\begin{equation}
{\cal M}\lbrace{\bf u}(x)\vert\tilde{{\bf f} }(x)\rbrace =\int{\cal
M}\lbrace{\bf u}'(x)\rbrace\delta \big[{\cal N}\lbrace{\bf u}'(x)\rbrace
-\tilde{{\bf f} }(x)\big]
D{\bf u}'(x)\  .
\label{20i13}
\end{equation}
Next we need to consider the average of ${\cal M}$ with respect to the
realizations of ${\bf u}$, (which in turn are determined by the
realizations of ${\tilde{\bf f}}$). As in the Wyld approach, it will be
assumed here that ${\tilde{\bf f} }$ has Gaussian statistics. The average
of any functional ${\cal P}\lbrace{\tilde{\bf f} }\rbrace$ can be written
as the Gaussian functional integral
\begin{eqnarray}
&&\langle {\cal P}\lbrace\tilde{{\bf f} }\rbrace\rangle ={1\over Z_1}
\int D\tilde{{\bf f} }(x){\cal P}
\lbrace{\tilde{\bf f} }\rbrace
\label{20i14} \\
&\times& \exp \Big[ -{1\over 2}\int \tilde f_\alpha
(x)D_{\alpha\beta}^{-1}(x-y)\tilde f_\beta (y)dxdy\Big]\  .
\nonumber
\end{eqnarray}
The covariance matrix $D_{\alpha\beta}$ is determined by the condition
that for the case ${\cal P}\lbrace\tilde{{\bf f} }\rbrace =\tilde f_\alpha
(x)\tilde f_\alpha (y)$ Eq.(\ref{20i14}) would lead to the equation
\begin{equation}
\langle\tilde f_\alpha (x)\tilde f_\alpha (y)\rangle
=D_{\alpha\beta}(x-y)\ .
\label{20i15}
\end{equation}
The partition sum $Z_1$ is
\begin{equation}Z_1=\int D\tilde{{\bf f}
}(x)\exp\Big[ -{1\over 2}\int\tilde f_\alpha
(x)D_{\alpha\beta}^{-1}(x-y)\tilde f_\beta (y)dxdy\Big] .
\label{20i16}
\end{equation}
In perfoming this average for the functional ${\cal M}\lbrace{\bf
u}(x)\vert\tilde{\bf f}(x)\rbrace$ of (\ref{20i13}) it is convenient to
rewrite the delta functional first in an exponential form such that all
the $\tilde f$ factors appear in the exponential.  We use the
representation of the 1-dimensional delta function
\begin{equation}
\delta (z)=\int^\infty_{-\infty}{dp\over  2\pi}\exp(-ipz)
\label{20i17}
\end{equation}
at every point of the grid.  Thus
\begin{eqnarray}
&&\delta \Big[{\cal N}\lbrace{\bf u}(x)\rbrace -{\tilde{\bf f} }(x)\Big]
=\lim_{M\rightarrow\infty}{1\over (2\pi)^M}\int\Big[ \prod^M_{i=1}dp_\alpha
(x_i)\Big]
\nonumber \\
&&\exp \Big\{ i\sum^M_{i=1} p_\alpha (x_i)\Big[ N_\alpha\lbrace{\bf
u}(x_i)\rbrace -\tilde f_\alpha (x_i)\Big]  \Big\}
\label{20i18}
\end{eqnarray}

We average now Eq.(\ref{20i13}) using the general Gaussian recipe
(\ref{20i14}), representing the delta function as the continuum limit of
(\ref{20i17}). The result reads
\begin{eqnarray}
&&\left\langle{\cal M}\lbrace{\bf u}(x)\rbrace\right \rangle =
{1\over Z_2}\int D{ \tilde{\bf f} }(x)D{\bf u}'(x)D{\bf p}{\cal M}
\lbrace{\bf  u}'(x)\rbrace
\nonumber\\
&\times&\exp\left\{ i\int dx {\bf p}(x)\cdot\Big[ {\cal N}\lbrace{\bf
u}'(x)\rbrace -\tilde{{\bf f} }(x)\Big] \right\}
\label{20i19}  \\
&\times&\exp\left\{ -{1\over 2}\int\tilde f_\alpha (x)
D_{\alpha\beta}^{-1}(x-y)\tilde f_\beta (y)dxdy)\right\}\ .
\nonumber
\end{eqnarray}
We remind the reader that our path integrals are limited to divergenceless
contributions, cf. (\ref{20i6}). In Eq. (\ref{20i19})  $Z_2$ is the
partition sum which is the RHS of (\ref{20i19}) without ${\cal
M}\lbrace{\bf u}(x)\rbrace$. Note that it serves to cancel the formally
divergent factor $(2\pi )^M$. Also note that  ${\bf p}$ appears only in
scalar products with divergenceless fields, and any projection onto vector
fields that are not divergenceless cancels from $D{\bf p}$ with the same
$D{\bf p}$ that appears in $Z_2$.  It will be convenient to take ${\bf p}$
to be divergenceless from the start, by defining $D{\bf p}$ similarly to
$D{\bf u}$ in Eq.(\ref{20i6}).

We can perform now the Gaussian integration over $D{\tilde{\bf f} }$.
One way of doing it is to change the ${\tilde{\bf f} }$ variables using a
similarity transformation that diagonalizes the matrix $D$, then to
perform the Gaussian integral at each space-time point independently, and
lastly to rotate back with the similarity transformation. The final
result is
\begin{eqnarray}
&&\langle{\cal M}\lbrace{\bf u}(x)\rbrace\rangle
={1\over Z}\int D{\bf u}'(x)D{\bf p} {\cal M}\lbrace{\bf u}'(x)\rbrace
\label{20i20} \\
&\times&\exp\Big\{ i\int p_\alpha(x){\cal N}_\alpha\{ {\bf u}'(x)\}dx
\nonumber\\ &-&{1\over 2}\int p_\alpha (x)D_{\alpha\beta}(x-y)p_\beta
(y)dxdy\Big\}\ .
\nonumber
\end{eqnarray}
It is quite evident that (\ref{20i20}) furnishes a starting point for the
calculation of any correlation function  $\langle{\bf u}(x_1){\bf
u}(x_2)\cdot\cdot\cdot{\bf u}(x_n)\rangle$. We recall  however that the
theory calls also for the calculation of the response  or the Green's
function. We show now that this calculation is also  available from
Eq.(\ref{20i20}). To see this imagine that we add to the force
${\tilde{\bf f} }$ some external deterministic component ${\bf h}( x)$ that
makes $\langle{\bf u}(x_1)\rangle$ non-zero. The addition of this
component changes $N_\alpha\{{\bf u}(x)\}$ in the exponent in
(\ref{20i20}) to $N_\alpha\lbrace{\bf u}(x)\rbrace -{\bf h}( x)$. Consider
now the response (\ref{20d2}) which in this case can be represented as
\begin{equation}
G_{\alpha\beta}(x_1,x_2)=i{\delta\langle u_\alpha (x_1)\rangle\over\delta
h_\beta (x_2)}\Bigg|_{h\rightarrow 0}\ .
\label{20i21}
\end{equation}
For a finite ${\bf h}$
\begin{eqnarray}
&&\langle u_\alpha (x_1)\rangle
={1\over Z}\int D{\bf u}(x)D{\bf p} u_\alpha (x_1)\exp\Big\{
i\int p_\beta[{\cal N}_\beta \lbrace{\bf u}(x)\rbrace
\nonumber \\
&-& h_\beta (x)]  dx -{1\over 2}\int p_\alpha
(x)D_{\alpha\beta} (x-y)p_\beta (y)dxdy\Big\}\ .
\label{20i22}
\end{eqnarray}
Using this we can compute (\ref{20i21}):
\begin{eqnarray}
&&G_{\alpha\beta} (x_1,x_2)  =
{1\over Z}\int D{\bf u}(x)D{\bf p} u_\alpha
(x_1)p_\beta (x_2)
\label{20i23} \\
&\times&\exp\lbrace i\int~p_\beta (x)[  {\cal
N}_\beta\lbrace{\bf u}(x)\rbrace] dx \nonumber\\ & -&{1\over
2}\int~p_\alpha (x)D_{\alpha\beta} (x-y)p_\beta (y)dxdy\rbrace =\langle
u_\alpha (x_1)p_\beta (x_2)\rangle \ .
\nonumber
\end{eqnarray}
 In the same way one can compute non-linear response functions like
\begin{eqnarray}
&&G_{_{\rm NL}}(x_1,x_2,x_3)
\label{20i24}   \\
&\equiv & -{\delta^2\langle u_\alpha
(x_1)\rangle\over \delta h_\beta (x_2)\delta h_\gamma
(x_2)}\Big\vert_{h\rightarrow 0}
=\langle u_\alpha (x_1)p_\beta (x_2)p_\gamma
(x_3)\rangle
\nonumber
\end{eqnarray}
etc. It is obvious that every additional functional derivative will appear
as another factor of ${\bf p}$ in the correlator. In general it is useful
to introduce the generating functional
\begin{equation}
Z({\bf l},{\bf m})\equiv\left\langle \exp\int dx[ {\bf u}(x)\cdot{\bf
l}(x) +{\bf p}(x)\cdot{\bf m}(x)] \right\rangle\ .  \label{20i25}
\end{equation}
All the needed statistical averages can be obtained as a functional
derivative of $Z({\bf l},{\bf m})$, taken at ${\bf l}={\bf m}=0$, for
example
 \begin{eqnarray}
 \langle u_\alpha (x_1)u_\beta (x_2)\rangle
&=&{\delta^2Z({\bf l},{\bf m})\over\delta l_\alpha (x_2)\delta l_\beta
(x_2)}
\label{20i26}\\
\langle u_\alpha (x_1)u_\beta (x_2)u_\gamma (x_3)\rangle
&=&{\delta^3Z({\bf l},{\bf m})\over\delta l_\alpha (x_2)\delta l_\beta
(x_2)\delta l_\gamma (x_3)}\,,
\label{20i27}
\end{eqnarray}
etc. Similarly one can compute any kind of Green's function. For example,
\begin{eqnarray}
G_{\alpha\beta}(x_1,x_2)
&=&{\delta^2Z({\bf l},{\bf m})\over\delta l_\alpha (x_2)\delta m_\beta
(x_2)}\,,
\label{20i28}\\
G_{_{\rm NL}} (x_1,x_2,x_3)
&=&{\delta^3Z({\bf l},{\bf m})\over\delta l_\alpha (x_2)\delta l_\beta
(x_2)\delta m_\gamma (x_3)}\,,
\label{20i29}\\
i{\delta\langle u_\alpha (x_1)u_\beta (x_2)\rangle\over
\delta h_\gamma (x_3)}
&=&{\delta ^3Z({\bf l},{\bf m})\over\delta l_\alpha (x_2)\delta l_\beta
(x_2)\delta m_\gamma (x_3)}\,,
\label{20i30}
\end{eqnarray}
etc. Finally we express the generating functional $Z({\bf l},{\bf m})$,
with the help of Eq.(\ref{20i20}), in the form of a functional integral:
\begin{eqnarray}
&& Z({\bf l},{\bf m})={1\over Z}\int D{\bf u}(x)D{\bf p(x)}
\label{20i31}\\
&\times& \exp\Big\{ iI +
\int dx[ {\bf u}(x)\cdot{\bf l}(x)+{\bf p}(x)\cdot{\bf m}(x)] \Big\}\ .
\nonumber
\end{eqnarray}
The quantity $I$ is referred to as the effective action, and for future
work it is useful to divide into two parts, the one quadratic and the
other triadic in the field ${\bf p}$ and ${\bf u}$:
\begin{eqnarray}
I&=&I_0+I_{\rm int}\,,
\label{20i32}\\
I_0&=&\int dx \Big[  p_\alpha {  \partial u_\alpha \over \partial t}
-\nu\, p_\alpha{\nabla}^2u_\alpha\Big]
\nonumber \\
&+&{i\over 2}\int p_\alpha (x)D_{\alpha\beta} (x-y)p_\beta (y)dxdy\,,
\label{20i33}\\
 I_{\rm int} & =& -i\int dx~p_\alpha (x)u_\beta (x)\partial_\beta
u_\alpha (x)
\label{20i34} \\
& =&  {1\over 2}    \int {dq_1dq_2dq_3\over (2\pi )^{12}}p_\alpha
\Gamma_{\alpha\beta\gamma}(q,q_1,q_2)u_\beta (q_2)u_\gamma (q_3)
\nonumber
\end{eqnarray}
The last line follows from the definition of the vertex in the $q$
representation, see Eq. (\ref{20b22}). Note that we did not display a
transverse projector in the expression of $I_{\rm int}$. The reason is
that the definition (\ref{20i6}) restricts anyway the integration to
divergenceless fields.

Finally, we introduce also the bare generating functional which is
(\ref{20i31}) when $I=I_0$,
\begin{eqnarray}
&&
Z_0({\bf l},{\bf m})={1\over Z_0}\int D{\bf u}(x)D{\bf p}
\label{20i35}   \\
&\times& \exp\Big\{ iI_0  +\int dx[ {\bf u}(x)\cdot {\bf I}(x)
+{\bf p}(x)\cdot{\bf m}(x)] \Big\}
\nonumber
\end{eqnarray}
This bare generating functional is used in the formulation of the
perturbative expansion. We will first use it to evaluate the bare
propagators.

\subsection{The evaluation of the bare propagators}
\label{20sect:pathint-C}
It is convenient to change the variables in $I_0$ from ${\bf p}(x)$ and
${\bf u}(x)$ to ${\bf p}({\bf k},\omega )$ and ${\bf u}({\bf k},\omega )$,
which are defined according to Eq. (\ref{20b19}).  The Jacobian of the
transformation (which is formally a divergent  constant) cancels with the
partition sum:
\begin{eqnarray}
I_0&=&
\int {dq\over (2\pi )^4}p_\alpha (q)[  -i\omega
+\nu k^2] u_\alpha (-q)
\nonumber\\
&+&
{1\over 2}\int {dq\over (2\pi )^4}p_\alpha (q)D_{\alpha\beta} (q)p_\beta
(-q)
\label{20j1}
\end{eqnarray}
where we remind the reader that $q\equiv ({\bf k},\omega )$.

The bare propagators are computed from $Z_0({\bf l},{\bf m})$ which in
this presentation is written as
\begin{eqnarray}
&&Z_0({\bf l},{\bf m}) ={1\over Z_0}\int D{\bf
u}(q)D{\bf p}(q)
\label{20j2} \\
&\times& \exp \Big\{ iI_0
+\int{dq\over (2\pi )^4}[ {\bf u}(q)\cdot {\bf l} (-q)+{\bf p}(q)\cdot{\bf
m}(-q)] \Big\}    \ .
\nonumber
\end{eqnarray}

This functional integral can be computed explicitly. Since we have no
mixture of different $q$'s in the exponent, the exponential can be computed
(on the grid) as the product of exponentials. The functional integration
is represented as a product of integrals, each one for one discrete $q$
and $-q$:
$$\int D{\bf u}(q)=\prod_{q>0}\int du(q)du(-q)\ .$$
Computing the resulting  Gaussian integrals, and using the expression of
the bare Green's function (\ref{20b16}), we end up with
\begin{eqnarray}
&&Z_0({\bf l},{\bf m})
\label{20j3}  \\
&=&\exp\Big\{ \int{dq\over (2\pi )^4}
\Big[ {1\over 2}l_\alpha   (-q)G^0_{\alpha\beta}(q)
D_{\beta\gamma}(q)G^{0*}_{\gamma\delta}(q)l_\delta (-q)
\nonumber \\
&+&G^0_{\alpha\beta} (q)l_\alpha (-q)m_\beta (-q)\Big] \Big\}\ .
\nonumber
\end{eqnarray}
 We can check that this zero$^{\rm th}$ order generating functional gives
 the same results as the zero$^{\rm th}$ order quantities defined in the
 direct perturbation theory. In computing the statistical quantities from
(\ref{20j3}) we use the general property of functional differentiation
\begin{equation}
{\delta l_\alpha (q)\over\delta l_\beta (q')}=(2\pi )^4\delta
(q-q')\delta_{\alpha\beta}\ .
\label{20j4}
\end{equation}
and the symmetry of the bare Green's function
$G^0_{\alpha\beta}(q)=-G^{0*}_{\alpha\beta}(-q)$ [cf. Eq.(\ref{20b16})].
For example, defining the velocity correlation function in
$q$-representation, $F_{\alpha\beta}(q)$, according to
 \begin{equation}
\langle u_\alpha (q)u_\beta (q')\rangle =(2\pi )^4\delta
(q+q')F_{\alpha\beta}(q)\,,
\label{20j5}
\end{equation}
we compute its bare value $\langle u_\alpha (q)u_\beta (q')\rangle^ 0$ as
\begin{equation}
\langle u_\alpha (q)u_\beta (q')\rangle^0 ={\delta^2Z_0({\bf l},{\bf
m})\over\delta l_\alpha (-q)\delta l_\beta
(-q')}\ .
\label{20j6}
\end{equation}
In other words, we find
\begin{equation}
F^0_{\alpha\beta}(q)=G^0_{\alpha\gamma}(q)D_{\gamma\delta}(q)G^*_
{\delta\beta}(q)\ .
\label{20j7}
\end{equation}
Similarly, defining the Green's function in $q$-representation as
\begin{eqnarray}
(2\pi )^4\delta (q&+&q')G_{\alpha\beta}(q)
\nonumber\\
&=&
i{\delta\langle u_\alpha (q) \rangle\over \delta h_\beta
(q')}\Bigg|_{h\rightarrow 0}=\langle u_\alpha (q)p_\beta
(q')\rangle\ .
\label{20j8}
\end{eqnarray}
Since
\begin{equation}
\langle u_\alpha (q)p_\beta (q')\rangle^0
={\delta ^2Z_0({\bf l},{\bf m})\over\delta m_\alpha (-q)\delta l_\beta
(-q')}\,,
\label{20j9}
\end{equation}
we get trivially that
$$\langle u_\alpha (q)p_\beta (q')\rangle^ 0=(2\pi
)^4\delta (q+q')G^0_{\alpha\beta} (q)\ .$$

We see that this theory generates the same zeroth order
propagators as the direct perturbation expansion of Section II. Note
however that there exists an apparent additional propagator here which is
absent in II, which is $\langle p_\alpha (q)p_\beta (q')\rangle$.
However, the zeroth order $\langle p_\alpha (q)p_\beta (q')\rangle^ 0$ is
zero. This follows from the fact that $Z_0({\bf l},{\bf m})$  does not
have a term quadratic in ${\bf m}$ in the exponent. We shall show later
that it is zero to all orders.

Next we show that the statistics inherited from $Z_0({\bf l},{\bf
m})$ are Gaussian statistics. The exponent contained in $Z_0({\bf
l},{\bf m})$ can be expanded as
\begin{eqnarray}
Z_0({\bf l},{\bf m}) &=& 1 +\int{dq\over (2\pi )^4}\Big[ {1\over
2}l_\alpha (-q)F^0_{\alpha\beta}(q)l_\beta (q)l_\beta (-q)
 \nonumber \\
&&\quad   +l_\alpha(-q)G^0_{\alpha\beta}(q)m_\beta (-q)\Big]
\label{20j10}  \\
&+&{1\over 2}\Bigg\{
\int{dq\over (2\pi )^4}\Big[ {1\over 2}l_\alpha
(-q)F^0_{\alpha\beta}(q)l_\beta (-q)
\nonumber \\
&+&l_\alpha
(-q)G^0_{\alpha\beta}(q)m_\beta (-q)\Big]\Bigg\}^2
+\dots
\nonumber \\
&+&{1\over n!}\Bigg\{
\int{dq\over (2\pi )^4}\Big[
{1\over 2}l_\alpha (-q)F^0_{\alpha\beta}(q)l_\beta (-q)
\nonumber \\
&+&l_\alpha (-q)G^0_{\alpha\beta}(q)m_\beta (-q)\Big]\Bigg\}^n +\dots
\nonumber
\end{eqnarray}
Clearly, if we take an odd number of derivatives with respect to
$l_\alpha$, and then send ${\bf l}$ and ${\bf m}$ to zero, the result
vanishes.  On the other hand, if we take $2n$ derivatives, and then send
${\bf l}$ and ${\bf m}$ to zero, only the contributions coming from the
$n$-th power of the integral survive. The answer will be proportional to
$[ F^0] ^n$ with $n$ delta functions.  The  number of terms will be $2n!/[
n!2^n]$  which is precisely the $(2n-1)!!$ number  of terms expected in
the Gaussian statistics (cf. Sect. II C).  Note that we have Gaussian
statistics also for the field ${\bf p}$.

Instead of looking at the correlation functions, one can study the
cumulants. These of course need to vanish beyond the lowest order.  An
immediate way to obtain the cumulants is to take functional derivatives
of $\ln\,Z_0({\bf l},{\bf m})$ instead of $Z_0({\bf l},{\bf m})$ itself.
Obviously,
\begin{eqnarray}
&& \ln Z_0({\bf l},{\bf m})=\int{dq\over (2\pi )^4}
\label{20j11}\\
&\times &\Big[ {1\over 2}l_\alpha (-q)F^0_{\alpha\beta}(q)l_\beta
(-q)+l_\alpha (-q)G^0_{\alpha\beta}(q)m_\beta (-q)]\  .
\nonumber
\end{eqnarray}
Only the second derivative with respect to ${\bf l}$ survives here,
giving us only the bare 2-point propagators $F^0_{\alpha\beta}$ and
$G^0_{\alpha\beta}$ as the non-vanishing  cumulants, as is required for
Gaussian statistics.

\begin{figure}
             \epsfxsize=8.6truecm
\centerline{             \epsfbox{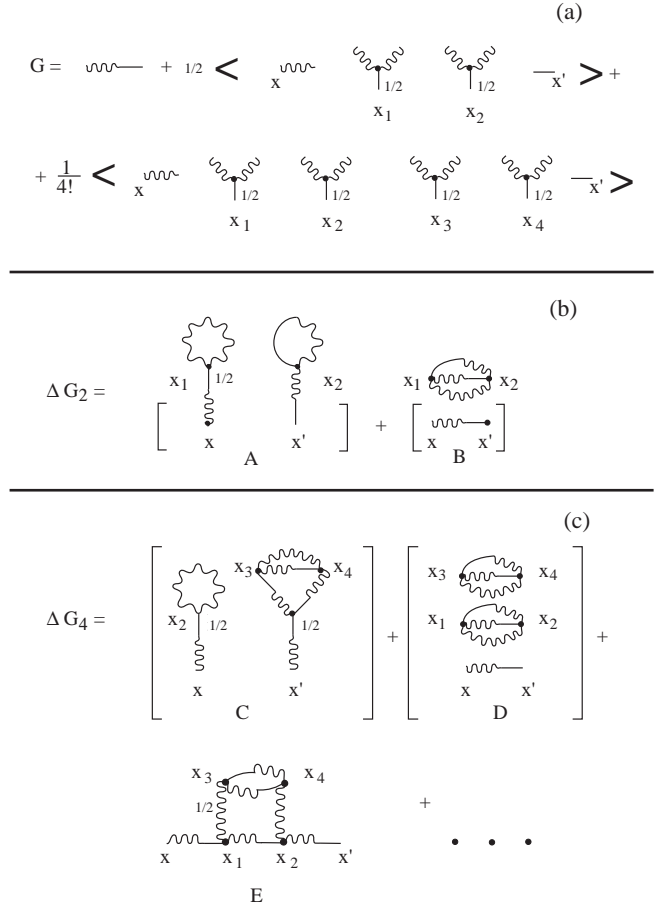} } 
             \vspace{.5cm}
\caption{
The graphic representation of Eq.(4.50). Panel (a): The bare, the
second  order and the fourth order terms. Panel (b): All the diagrams up
to order {$\Gamma ^2$}  and (c) representative diagrams of order {$\Gamma
^4$} that did not appear in the  Wyld diagrammatic series in
Fig.~5. All these new diagrams vanish.       }
             \label{20-fig12}
             \end{figure}

\subsection{Diagrammatic expansion}
\label{20sect:pathint-D}
 We return to the generating functional $Z({\bf l},{\bf m})$ of
 Eq.(\ref{20i31}) and write now the propagator $F_{\alpha\beta}$ and
$G_{\alpha\beta}$ as
\begin{eqnarray}
F_{\alpha\beta}(x,x')&=&
\langle u_\alpha (x)u_\beta (x')\rangle
\label{20k1}\\
&=&\langle u_\alpha (x)u_\beta
(x')\exp \{iI_{\rm int}\}\rangle^0\,,
\nonumber\\
G_{\alpha\beta}(x,x')&=&
\langle u_\alpha (x)p_\beta (x')\rangle
\label{20k2}\\
&=&\langle  u_\alpha (x)p_\beta  (x')\exp\{iI_{\rm int}\}\rangle^0\,,
\nonumber
\end{eqnarray}
with $iI_{\rm int}$ as defined as [cf. (\ref{20b18}) and (\ref{20i34})]
\begin{equation}
iI_{\rm int}={1\over 2}\int dx~p_\alpha
(x)\Big\{ \Gamma^{\cdot {\bf u}(x)}_{\cdot {\bf
u}(x)}\Big\}_\alpha
\label{20k3}
\end{equation}
Expanding the exponential in (\ref{20k1}) and (\ref{20k2}) we find
\begin{figure}
 \epsfxsize=8.6truecm
\centerline{ \epsfbox{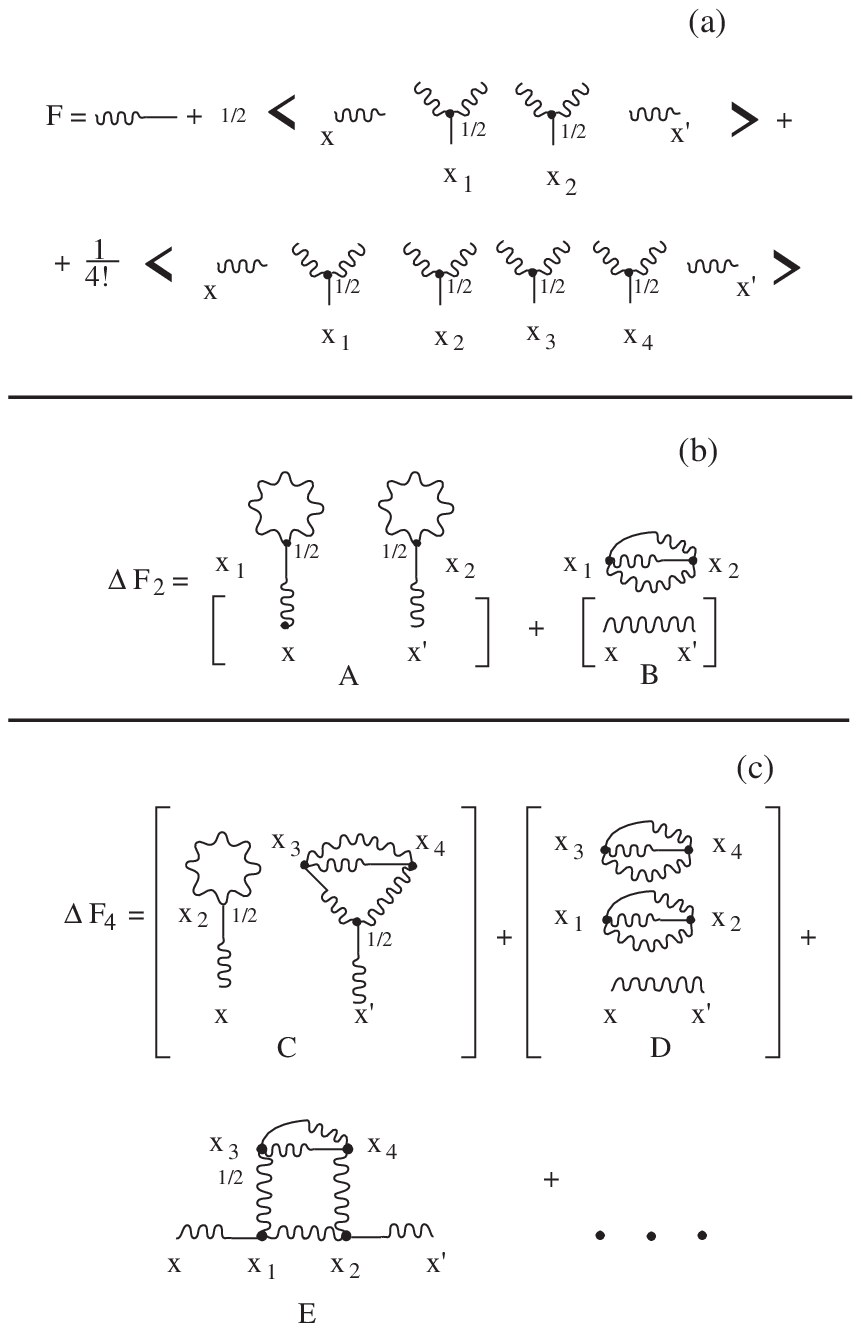} } 
  \vspace{.5cm}
\caption{The graphic representation of Eq.(4.51). Panel (a): The bare,
the second  order and the fourth order terms. Panel (b): All the diagrams
up to order {$\Gamma ^2$} and representative diagrams of order
{$\Gamma ^4$} that did not appear in the  Wyld diagrammatic series in
Fig.~6. All   these new diagrams vanish.  }
\label{20-fig13}
\end{figure}

\begin{eqnarray}
G_{\alpha\beta}(x,x')&=&
\sum_n{1\over n!}\Big\langle
u_\alpha (x)p_\beta (x')
\nonumber\\
&\times&\Big\{
{1\over 2}\int dx~p_\alpha (x)\Big[
\Gamma^{\cdot {\bf u}(x)}_{\cdot  {\bf
u}(x)}\Big]_\alpha\Big\}^n\Big\rangle^0\,,
\label{20k4}\\
F_{\alpha\beta}(x,x)&=&
\sum_n{1\over n!}\Big\langle u_\alpha (x)p_\beta (x)
\nonumber\\
&\times& \Big\{ {1\over 2}\int dx~p_\alpha (x)\Big[
\Gamma^{\cdot {\bf u}(x)}_{\cdot {\bf
u}(x)}\Big]_\alpha\Big\}^n\Big\rangle^0\,,
\label{20k5}
\end{eqnarray}
Such an expansion contains products of the fields ${\bf u}$ and ${\bf p}$
averaged over the Gaussian ensemble. As explained above, only products of
$\langle u_\alpha (x)p_\beta (x')\rangle^0$ and $\langle u_\alpha
(x)u_\beta (x')\rangle^0$ can appear in such a series.  (Remember that
$\langle p_\alpha (x)p_\beta (x')\rangle^0=0)$. As done in chapter II, the
procedure of computing the averages can be represented  diagrammatically
by pairing all the allowed pairs of fields and gluing.  The needed graphic
notation is the same as the one shown in Fig.~\ref{20-fig1}. The  field
${\bf u}$ is represented by a short wavy line, the field ${\bf p}$ by a
short straight line, and now it is obvious why the double correlator is
represented by a twice long wavy line. Also, the notation for the  Green's
function becomes natural now, being a correlator of ${\bf u}$ and ${\bf
p}$.  In addition, the notation for the vertex as a dot connecting one
straight  and two wavy tails becomes apparent in Eqs.(\ref{20k4}) and
(\ref{20k5}).

The diagrams that are obtained from Eqs.(\ref{20k4}) and (\ref{20k5}) are
shown in Figs.~\ref{20-fig12}a and ~\ref{20-fig13}a respectively. Upon
pairing all the possible pairs and gluing, we get exactly the diagrams
appearing in Figs.~\ref{20-fig5}a and ~\ref{20-fig6}a respectively, and
in addition we get the diagrams appearing in panels b and c of
Figs.~\ref{20-fig12} and ~\ref{20-fig13}. These additional diagrams,
that are denoted here as {$\Delta G_2,\  \Delta G_4,\  \Delta F_2$} and
{$\Delta F_4$} respectively, do not appear in the diagrammatic expansion
leading to Figs.~\ref{20-fig5} and ~\ref{20-fig6}. However, it is  easy
to see that they are all zero. The diagram A in ~\ref{20-fig12}b
contains one term which is a diagram for
{$\langle{\bf u}\rangle$} and another one  which represents a contribution
to {$\langle{\bf p}\rangle$}.  The latter contains a frequency integral
over {$G^0$}, which is $-i/2$.  However the vertex {$\Gamma$} in front
of this contains a factor of {${\bf k} $} which is odd in the {${\bf
k}$}-integration. Therefore the diagram vanishes.  The diagrams B, C, D
and E in ~\ref{20-fig12}b,c contain closed loops of Green's functions.
Such loops vanish because of causality: in $t$-representation only
{$t=0$} is allowed for the Green's function in a closed loop, and thus the
integral over time has a contribution only from a single point, and it
vanishes.

Next we argue that this phenomenon is general: all the diagrams that do
not appear in Wyld's expansion are zero. These diagrams are either
unlinked (like all the diagrams in Fig.~\ref{20-fig12}  except for E) or
linked to the body via wavy lines (like E). All these diagrams must have a
closed loop of Green's functions. Consider a diagram from the $n$-th
order, in which $m$ vertices, $m\leq n$, form a fragment that is unlinked
to the body. Glue the straight line of the first vertex to the wavy line
of another. Then either we close the straight line of the second vertex on
the wavy line, or we connect it to a wavy line of a third vertex. The
first possibility leads to zero due to the argument of diagram A in
{}~\ref{20-fig12}b.  In the second possibility we again need to repeat the
same option, or to closed with the first vertex. The production of a close
loop of Green's functions is inevitable, and these vanish due to
causality. Diagrams that are connected only via wavy lines also must have
a closed loop of Green's functions for the same reasons.

We thus conclude that the path integral formulation leads to exactly the
same diagrams as the Wyld direct perturbative expansion.  This fact is not
widely recognized. One reason for this is that in Ref.\cite{61Wyl} the
three types of vertices which appear in this theory were not explictly
discussed. They were first dealt with properly in \cite{73MSR} in which it
was stated that the results agree with \cite{61Wyl} only to order
$\Gamma^3$. We have shown above that the two techniques generate
exactly the same (non-zero) diagrams on the naive level, and these can be
resummed in exactly the same way. For example, Fig.~\ref{20-fig6} of
Ref\cite{93LL} was claimed to differ from the Wyld technique, but in
fact is identical to the present Fig.~\ref{20-fig10}b which was obtained
entirely within the Wyld expansion.

\section{Suggestions for further reading}

The material presented in these notes should be sufficient to prepare the
student for reading the research papers on the subject of renormalized
perturbation theory of turbulence. For the convenience of the interested
student we present in this section a brief guided tour through the
latest publications of our own group.
The next paper to read is \cite{95LP-b} in which the Belinicher-L'vov
renormalization scheme is used to prove that the theory for the
correlation functions and the Green's functions is finite. In
other words, it is shown (order by order) that for Re$\to\infty$, after
the appropriate resummations, all the diagrams converge both in the IR and
the UV limits.  This fact tells us that there is no length scale available
to form dimensionless corrections to the K41 scaling. In \cite{95LP-c} the
issue of anomalous scaling is discussed. It is shown that there exist
anomalous exponents in turbulence, but they appear in gradient fields and
their correlations. Thus for example the dissipation field is anomalous,
and its correlation function is strongly renormalized compared with the
Gaussian limit. The issue of the effect of the anomalous fields on the
structure functions was discussed in \cite{95LP}. It is argued there that
the anomalous fields are subcritical, and that they influence the
structure functions only on the level of corrections to scaling. The
latter corrections are large, and they disappear only at very large
values of Re. We attributed the observed deviations in experiments to
these ``subciritcal" corrections. Two more papers that may be considered
are \cite{95LP-f} and \cite{95LP-g} in which further relations to
experiments are discussed. In \cite{95LP-g} we discuss which length scale
should be used as the renormalization scale in turbulence, i.e the inner
scale $\eta$ or the outer scale $L$. In \cite{95LP-f} one can find an
experimental test of some aspects of the new theory, especially some
unusual correlation functions between velocity differences and the
dissipation field.

\acknowledgments
We thank Adrienne Fairhall and Volodia Lebedev for their comments on the
manuscript.  This work has been supported in part by the German-Israeli
Foundation and the N. Bronicki fund.

\section{Appendix:~~The Jacobian}
In this appendix we explain the known fact\cite{76Dom,76Jan} that the
Jacobian that appears in Eq.(4.10) is unity.  From the definition of the
delta function it is obvious that

\begin{equation}
\int D {\bf u}(x)\delta\big[ {\bf u}(x) -{\cal N}^{-1}\{ \tilde{\bf f}(x)
\} \big] =1 \ .
\label{A1}
\end{equation}
This means that also the functional integral over the RHS of Eq.(4.10)
is unity. What we want to prove now is that also the functional integral
on the RHS of (4.10) without $J\{ {\bf u} \}$ is unity:
\begin{equation}
\int D {\bf u}(x)\delta\big[ {\cal N} \{ {\bf u}(x)  \}  - \tilde{\bf
f}(x) \big] =1  \ .
\label{A2}
\end{equation}
This will be the first indication that the Jacobian $J\{ {\bf u} \}=1$.
To accomplish this we will only use the fact that the Navier-Stokes
equations are first order in the time derivative:
\begin{equation} {\cal N} \{ {\bf u}(x) \} ={\partial {\bf u}(x)\over
\partial t} + {\rm NL}  \{ {\bf u}(x) \} \ .
\label{A3}
\end{equation}
The time derivative on the time grid may be represented in the retarded
convention
\begin{equation}
{\partial {\bf u}_n({\bf r})\over \partial t_n}
=  \big[ {\bf u}_n({\bf r})- {\bf u}_{n-1} ({\bf r})\big] /\tau  \ .
\label{A4}
\end{equation}
Using this representation we rewrite (\ref{A2}) on the time grid as
\begin{equation}
\int \prod_n D  {\bf u}_n({\bf r})  \delta \big[ {\bf u}_n({\bf r})
+  \widetilde{\rm NL}  \{ {\bf u}_{n-1}({\bf r})  \}   \big] \ .
\label{A5}
\end{equation}
We have collected everything except ${\bf u}_n$ itself into an operator
that was denoted as $\widetilde{\rm NL}$. It is obvious now that every
$D{\bf u}_n({\bf r})$ integration yields unity due to the delta function,
and finally the result is unity.  Next one should prove that the
functional integrals (\ref{A1}) and (\ref{A2}) but with any arbitrary
weight functional of $\bf u$ are also the same. Repeating the same
considerations on the time grid furnishes this demonstration with
essentially the same ease.


\begin{thebibliography}{10}

\bibitem[\ast]{lvov}
L'vov's e-mail: fnlvov@wis.weizmann.ac.il.

\bibitem[\dag]{procaccia}
Procaccia's e-mail: cfprocac@weizmann.weizmann.ac.il.

\bibitem{22Ric}
L.F. Richardson.
\newblock {\em Weather Prediction by Numeric Process}.
\newblock Cambridge Univ. Press, Cambridge, 1922.

\bibitem{26Ric}
Lewis~F. Richardson.
\newblock Atmospheric diffusion shown on a distance neighbor graph.
\newblock {\em Proc. R. Soc. Lond. A}, 110:709--737, 1926.

\bibitem{MY-2}
A.~S. Monin and A.~M. Yaglom.
\newblock {\em Statistical Fluid Mechanics: Mechanics of Turbulence},
  volume~II.
\newblock The MIT Press, Cambridge, Mass., 1973.

\bibitem{41Kol-a}
A.~N. Kolmogoroff.
\newblock \"uber das logarithmisch normale {V}erteilungsgesetz dd {D}imensionen
  der {T}eilchen bei {Z}erztckelungg.
\newblock {\em C.~R.~(Doklady) Acad.~Sc. URSS}, 31:99--101, 1941.

\bibitem{Fri}
Uriel Frisch.
\newblock {\em Turbulence: The Legacy of A.N. Kolmogorov}.
\newblock Cambridge University Press, Cambridge, 1995.
\newblock In press.

\bibitem{93SK}
K.R. Sreenivasan and P.~Kailasnath.
\newblock An update on the intermittency exponent in turbulence.
\newblock {\em Phys.~Fluids~A}, 5(2):512--514, 1993.

\bibitem{59Kra}
R.H. Kraichnan.
\newblock The structure of isotropic turbulence at very high {R}eynolds
  numbers.
\newblock {\em J.~Fluid~Mech.}, 5:497--543, 1959.

\bibitem{70Ors}
S.~A. Orszag.
\newblock Analytical theories of turbulence.
\newblock {\em J.~Fluid~Mech.}, 41:363--386, 1970.

\bibitem{61Wyl}
H.~W. Wyld.
\newblock Formulation of the theory of turbulence in an incompressible fluid.
\newblock {\em Ann.~Phys.}, 14:143--165, 1961.

\bibitem{73MSR}
P.~C. Martin, E.~D. Siggia, and H.~A. Rose.
\newblock Statistical dynamics of classical systems.
\newblock {\em Phys.~Rev.~A}, 8(1):423--437, July 1973.

\bibitem{75ZL}
V.~E. Zakharov and V.~S. L'vov.
\newblock On statistical description of the nonlinear wave fields.
\newblock {\em Quan. Electronics}, 18(10):1084--1097, 1975.

\bibitem{65Kra}
R.~H. Kraichnan.
\newblock Lagrangian-history closure approximation for turbulence.
\newblock {\em Phys.~Fluids}, 8:575--598, 1965.

\bibitem{66Kra}
R.H. Kraichnan.
\newblock Isotropic turbulence and inertial-range structure.
\newblock {\em Phys.~Fluids}, 9:1728--1752.

\bibitem{87BL}
V.~I. Belinicher and V.~S. L'vov.
\newblock A scale-invariant theory of fully developed hydrodynamic turbulence.
\newblock {\em Sov.~Phys.~JETP}, 66:303--313, 1987.

\bibitem{91Lvo}
V.~S. L'vov.
\newblock Scale invariant theory of fully developed turbulence. {H}amiltonian
  approach.
\newblock {\em Phys.~Rep.}, 207(1):1--47, August 1991.

\bibitem{95LP-b}
V.~S. L'vov and I.~Procaccia.
\newblock Exact resummations in the theory of hydrodynamic turbulence. {I}.
  {T}he ball of locality and normal scaling.
\newblock {\em Phys.~Rev.~E}, 1995.
\newblock Submitted.

\bibitem{76Dom}
C.~DeDominics.
\newblock {\em J.~Physique (Paris)}, 37:C1--247, 1976.

\bibitem{76Jan}
H.K. Jansen.
\newblock {\em Z.~Phys.~B}, 23:377, 1976.

\bibitem{95LP-c}
V.~S. L'vov and I.~Procaccia.
\newblock Exact resummations in the theory of hydrodynamic turbulence. {II}.
  {A} ladder to anomalous scaling.
\newblock {\em Phys.~Rev.~E}, 1995.
\newblock Submitted.

\bibitem{95LP}
V.~S. L'vov and I.~Procaccia.
\newblock ``{I}ntermittency'' in turbulence as intermediate asymptotics to
  {K}olmogorov'41 scaling.
\newblock {\em Phys.~Rev.~Lett.}, 1995.
\newblock Submitted, E-print in Los-Alamos e-board No 9410002 (1994)
  chao-dyn@xyz.land.gov~~~~.

\bibitem{93LL}
V.~S. L'vov and V.~V. Lebedev.
\newblock Exact relations in the theory of fully developed hydrodynamic
  turbulence.
\newblock {\em Phys.~Rev.~E}, 47(4):1794--1802, 1993.

\bibitem{95LP-f}
V.~S. L'vov and I.~Procaccia.
\newblock Correlator of velocity differences and energy dissipation as element
  in the subcritical scenario for non{K}olmogorov scaling in turbulence.
\newblock {\em Europhys. Lett.}, 1995.
\newblock In press.

\bibitem{95LP-g}
V.S. L'vov and I.~Procaccia.
\newblock Is the fundamental length scale in developed turbulence the
inner or the outer scale ?
\newblock {\em Phys.~Rev.~E}, 1995. Submitted

\end{thebibliography}

\end{document}